 \documentclass[aps,prx,twocolumn,superscriptaddress,nofootinbib,notitlepage,longbibliography]{revtex4-2}
\usepackage{amsfonts}
\usepackage{mathrsfs}
\usepackage{amsmath}
\usepackage{color}
\usepackage{natbib}
\usepackage{graphicx}
\usepackage{bm}
\usepackage{amssymb}
\usepackage{mathptmx}
\usepackage{stmaryrd}
\usepackage{xspace}
\usepackage{epstopdf}
\usepackage{dcolumn}
\usepackage{longtable}
\usepackage{multirow}
\usepackage{bbding}

\usepackage[colorlinks=true, letterpaper=true, pdfstartview=FitV, linkcolor=blue, citecolor=blue, urlcolor=blue]{hyperref}
\usepackage{gensymb}
\usepackage{subscript}

\makeatletter

\newcommand{\Rmnum}[1]{\expandafter\@slowromancap\romannumeral #1@}

\renewcommand\@biblabel[1]{#1.}
\makeatother

\begin{document}
	
	\title{Design monolayer iodinenes based on halogen bond and tiling theory}
	\author{Kejun Yu}
	\affiliation{Centre for Quantum Physics, Key Laboratory of Advanced Optoelectronic Quantum Architecture and Measurement (MOE), School of Physics, Beijing Institute of Technology, Beijing, 100081, China}
	\affiliation{Beijing Key Lab of Nanophotonics \& Ultrafine Optoelectronic Systems, School of Physics, Beijing Institute of Technology, Beijing, 100081, China}
	
	\author{Yichen Liu}
	\affiliation{Centre for Quantum Physics, Key Laboratory of Advanced Optoelectronic Quantum Architecture and Measurement (MOE), School of Physics, Beijing Institute of Technology, Beijing, 100081, China}
	\affiliation{Beijing Key Lab of Nanophotonics \& Ultrafine Optoelectronic Systems, School of Physics, Beijing Institute of Technology, Beijing, 100081, China}
	
	\author{Botao Fu}
	\affiliation{College of Physics and Electronic Engineering, Center for Computational Sciences, Sichuan Normal University, Chengdu, 610068, China}
	
	\author{Runwu Zhang}
	\affiliation{Centre for Quantum Physics, Key Laboratory of Advanced Optoelectronic Quantum Architecture and Measurement (MOE), School of Physics, Beijing Institute of Technology, Beijing, 100081, China}
	\affiliation{Beijing Key Lab of Nanophotonics \& Ultrafine Optoelectronic Systems, School of Physics, Beijing Institute of Technology, Beijing, 100081, China}
	
	\author{Da-shuai Ma}
	\affiliation{Institute for Structure and Function \& Department of Physics \& Chongqing Key Laboratory for Strongly Coupled Physics, Chongqing University, Chongqing 400044, P. R. China}
	\affiliation{Center of Quantum materials and devices, Chongqing University, Chongqing 400044, P. R. China}
	
	\author{Xiao-ping Li}
	\affiliation{School of Physical Science and Technology, Inner Mongolia University, Hohhot 010021, China}
	
	\author{Zhi-Ming Yu}
	\email{zhiming\_yu@bit.edu.cn}
	\affiliation{Centre for Quantum Physics, Key Laboratory of Advanced Optoelectronic Quantum Architecture and Measurement (MOE), School of Physics, Beijing Institute of Technology, Beijing, 100081, China}
	\affiliation{Beijing Key Lab of Nanophotonics \& Ultrafine Optoelectronic Systems, School of Physics, Beijing Institute of Technology, Beijing, 100081, China}
	
	\author{Cheng-Cheng Liu}
	\email{ccliu@bit.edu.cn}
	\affiliation{Centre for Quantum Physics, Key Laboratory of Advanced Optoelectronic Quantum Architecture and Measurement (MOE), School of Physics, Beijing Institute of Technology, Beijing, 100081, China}
	\affiliation{Beijing Key Lab of Nanophotonics \& Ultrafine Optoelectronic Systems, School of Physics, Beijing Institute of Technology, Beijing, 100081, China}
	
	\author{Yugui Yao}
	\affiliation{Centre for Quantum Physics, Key Laboratory of Advanced Optoelectronic Quantum Architecture and Measurement (MOE), School of Physics, Beijing Institute of Technology, Beijing, 100081, China}
	\affiliation{Beijing Key Lab of Nanophotonics \& Ultrafine Optoelectronic Systems, School of Physics, Beijing Institute of Technology, Beijing, 100081, China}
	
\begin{abstract}
	Xenes, two-dimensional (2D) monolayers composed of a single element, with graphene as a typical representative, have attracted widespread attention. Most of the previous Xenes, X from group-IIIA to group-VIA elements have bonding characteristics of covalent bonds. In this work, we for the first time unveil the pivotal role of a halogen bond, which is a distinctive type of bonding with interaction strength between that of a covalent bond and a van der Waals interaction, in 2D group-VIIA monolayers. Combing the ingenious non-edge-to-edge tiling theory and state-of-the-art ab initio method with refined local density functional M06-L, we provide a precise and effective bottom-up construction of 2D iodine monolayer sheets, iodinenes, primarily governed by halogen bonds, and successfully design a category of stable iodinenes, encompassing herringbone, Pythagorean, gyrated truncated hexagonal, i.e. diatomic-kagome, and gyrated hexagonal tiling pattern. These iodinene structures exhibit a wealth of properties, such as nontrivial novel topology, flat bands and fascinating optical characteristics, offering valuable insights and guidance for future experimental investigations. Our work not only unveils the unexplored halogen bonding mechanism in 2D materials but also opens a new avenue for designing other non-covalent bonding 2D materials.
\end{abstract}

\maketitle

\textit{Introduction. ---}
	Two-dimensional (2D) materials have been a hot topic since the discovery of graphene. Up to now, as one of the most important members of 2D materials, increasing Xenes made from group-III to VI elements have been predicted and synthesized, e.g., borophene~\cite{Synthesisborophenes,ZengXiaoCheng}, silicene~\cite{silicene2010}, phosphorene~\cite{Phosphorene2014} and tellurene~\cite{Te1Based1Monolayer}. Generally, disparate valence electron configurations among diverse main group elements engender markedly distinct structural arrangements, bonding behaviors, and material characteristics within the various Xene families. Yet, as one of the most intriguing members of the Xene family, rare study on group-VII Xenes has been done. Undoubtedly, exploring the theoretical foundations of group-VII Xenes will enhance our understanding of 2D materials and generate a more extensive range of practical applications.

	Solid halogen crystals exhibit a layered structure in the Cmca space group~\cite{solid1967iodine}, with stacked planar monolayers. Among the halogens, only iodine maintains its crystalline phase under ambient conditions. The halogen bond (XB, X stands for halogen), akin to hydrogen bonds as a non-covalent interaction~\cite{HOF1997JACS,HOF2019CSR}, plays a crucial role in stabilizing bulk iodine, despite often being overlooked. Recently, few-layer iodine nanosheets have been experimentally obtained from bulk iodine by physical exfoliation~\cite{FL1iodinene2020AM,FL1iodinene2021CPC,FL1iodinene2023JACS,FL1iodinene2023BS}, but it remains unclear about the fine structure of the iodinenes. On the other hand, it is noteworthy that XBs have been utilized for fabricating halogen-bonded organic frameworks (XOFs)~\cite{gong2021XOF1Angewandte,tetrabromobenzene} successfully with reliable stability in both solid and solution phases. Then the questions arise naturally: could 2D iodinenes be constructed by utilizing halogen bonding interactions, how to design them, and what interesting properties do such 2D materials have?

	In this work, based on the XBs, we predict a series of monolayer iodinenes as a new category of 2D materials utilizing the tiling theory and first-principle calculations. First of all, through our meticulous DFT calculations, including the calculations of lattice parameters and phonon spectra, we find that M06-L~\cite{M062006L} is an appropriate exchange-correlation functional for evaluating the XB and realize that the XB plays a key role in the formation of stable structures of halogen crystals. Herringbone iodinene, as monolayer iodinene exfoliated from bulk iodine, shows dynamic stability resulting from XB networks. Afterward, according to the interaction model of XB and taking into account the diversity of XB networks, we systematically design a series of structures of iodinenes based on halogen-bonded iodine molecules and the tiling theory to map the structure. Finally, we investigate the electronic structures of our designed iodinenes. All of those halogen-bonded iodinenes are semiconductors and exhibit nontrivial band topology, including herringbone, Pythagorean, gyrated truncated hexagonal (GTH), i.e. diatomic-kagome, and gyrated hexagonal (GH) iodinenes. Specifically, herringbone, Pythagorean, GTH/diatomic-kagome, and GH iodinenes are intrinsic Stiefel-Whitney topological insulators with higher-order bulk-edge correspondences, among which Pythagorean, GTH/diatomic-kagome, and GH iodinenes are $Z_2$ topological insulators exhibiting helical edge states with appropriate filling. These iodinenes possess flat bands that result from XB interactions and special structure, leading to remarkable absorption coefficients that exceed $ 10^{5}  cm^{-1} $ in the visible region, which reveals potential for photoelectronic application.
	
	\begin{figure}
		\includegraphics[width=8.5cm]{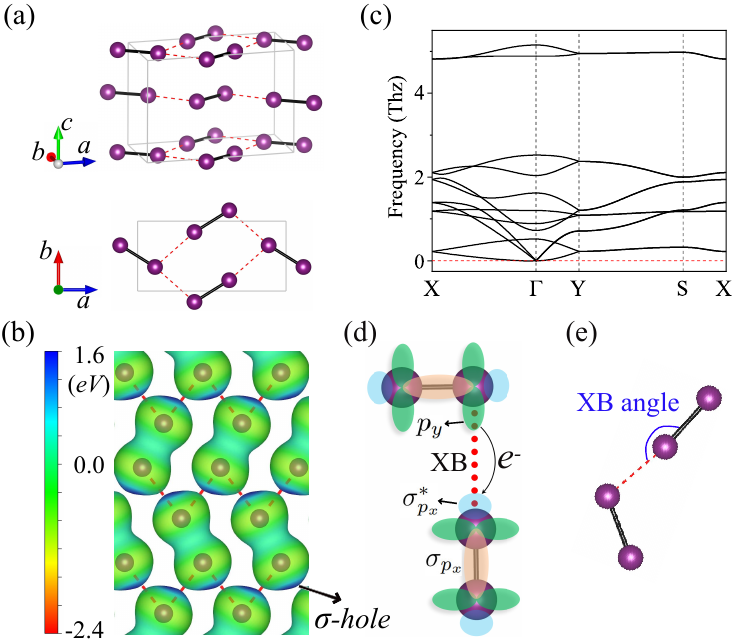}
		\caption{(a) 3D crystal structure (the upper panel) of bulk iodine and its monolayer (the lower panel). Black bold and red dashed lines indicate intramolecular covalent bonds and intermolecular halogen bonds (XBs), respectively. (b) Electrostatic potential (ESP) map on 0.015 a.u. charge density isosurface of monolayer herringbone iodinene. $\sigma$-hole occurs at the blue tip on the isosurface with positive electrostatic potential. (c) Phonon spectra of monolayer herringbone iodinene. (d) Schematic diagram of XB between two iodine molecules. The iodine atoms, $p_y$ orbitals (lone pairs), $\sigma_{p_x}$ bond and $\sigma^{*}_{p_x}$ antibond orbital ($\sigma$-hole) are depicted as purple balls, green spindles, yellow ellipsoid and blue ellipsoid, respectively. $p_z$ and $s$ orbitals are not shown here because $p_z$ sits out of the plane and $s$ is the core orbital. The red dotted line signifies the presence of the XB, involving the transfer of electrons from a lone pair to a $\sigma$-hole through a donor-acceptor interaction. (e) Schematic diagram of XB angle.}
		\label{fig1}
	\end{figure}

\textit{The halogen bond and selection of DFT functional. ---}
	Figure~\ref{fig1}(a) depicts the layered structure of bulk iodine. Specifically, each planar layer within the bulk iodine is formed by halogen-bonded iodine molecules. The electrostatic interaction model offers an intuitive image of XBs, as depicted in Fig.~\ref{fig1}(b), which displays the electrostatic potential (ESP) map of a monolayer iodine sheet. The unique ESP map is derived from the inherent anisotropic charge density distribution (see Fig. S1(b)) of iodine molecules themselves, where the electron density is accumulated around the equator and depleted on the elongation of the covalent bond. The depleted region is the so-called $\sigma$-hole~\cite{clark2007Sigmahole}, corresponding to the blue tip in Fig.~\ref{fig1}(b) with positive ESP. The XB is defined as the attraction between a nucleophilic region and the positive electrostatic region of a $\sigma$-hole, analogous to a hydrogen bond~\cite{XBdefinition}. Additionally, weak Van Der Waals (vdW) forces dominate the interaction between layers along the $\mathbf{c}$ axis in the bulk iodine. This arrangement reveals the potential for exfoliation to yield a freestanding monolayer of iodinene, given the significantly greater strength of the XB compared to vdW interactions.

	A comparative analysis of various GGA and meta-GGA functionals was conducted to evaluate their computational accuracy in modeling the interactions within bulk iodine. The results indicate that M06-L accurately captures the attractive XB interaction in bulk iodine, whereas SCAN~\cite{SCAN2015} does not exhibit this capability (see Fig. S2). This finding is consistent with the study by George et al.~\cite{Solid2015george}. It is pertinent to highlight that the utilization of PBE~\cite{PBE1996}  and PBEsol~\cite{PBEsol}  functionals for assessing the interaction associated with halogen bonding is deemed inappropriate (as delineated in Fig. S1\&S2 and Table S1), despite their application in recent research~\cite{Pytha1iodinene2021}. To account for long-range dispersive interactions, the DFT-D3 method~\cite{GrimmeDFTD3} was employed, resulting in a more accurate geometry configuration for bulk iodine (as demonstrated in Table. S1). Therefore, M06-L+D3 was chosen for the DFT calculations. Further computational details can be found in Supplemental Material (SM)~\cite{SM}.

	\begin{figure*}
		\includegraphics[width=16 cm]{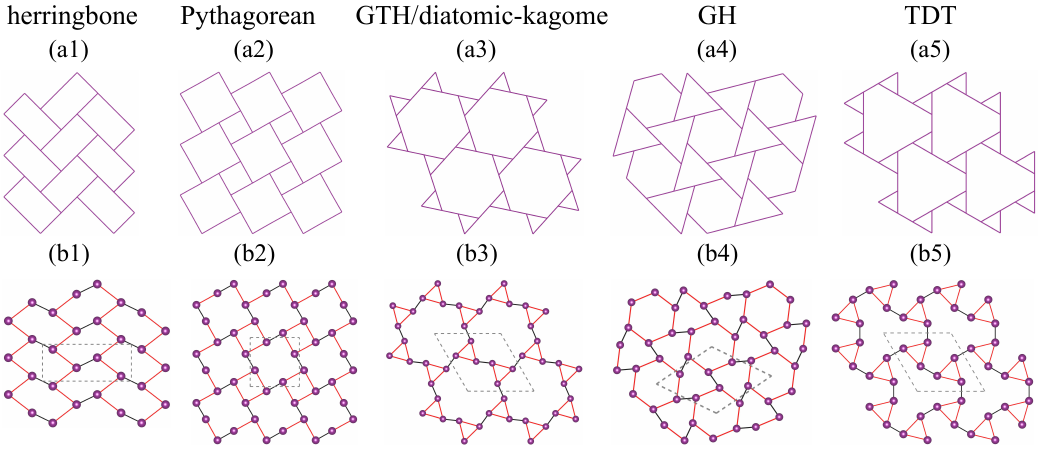}
		\caption{ (a1-a5) Five non-edge-to-edge tilings for mapping to desired monolayer iodinenes and (b1-b5) corresponding relaxed crystal structures. The nomenclature of those tilings is adopted in a visualized way. herringbone: herringbone tiling; GTH: gyrated truncated hexagonal tiling, which is topologically equivalent to diatomic-kagome structure; GH: gyrated hexagonal tiling; TDT: truncated distorted trihexagonal tiling. In (b1-b5), the covalent bond and halogen bond are depicted by black and red lines respectively. Grey dot lines indicate the unit cell.}
		\label{fig2}
	\end{figure*}

\textit{Design iodinenes with the tiling theory. ---} Firstly, a monolayer iodinene is exfoliated from bulk iodine, and we call it herringbone iodinene since the profile resembles herringbone. After relaxation, the intramolecular bond length and XB length are 2.75 and 3.60 {\AA} respectively with only minor deviations from the bulk counterparts. Additionally, the electrostatic potential (ESP) map of the herringbone iodinene is similar to that of the monolayer iodine sheet in bulk [see Fig.~\ref{fig1}(b)]. There is no imaginary frequency in phonon spectra (Fig.~\ref{fig1}(c)) of freestanding monolayer iodinene, revealing its dynamic stability. The XB network plays a crucial role in enabling individual iodine molecules to link with each other to form a planar structure, while herringbone iodinene would bear dynamic instability if XB is not included (see Fig. S5).
	
	Considering the crucial role of XB in forming stable bulk iodine and exfoliated monolayer, we now provide a general and effective bottom-up construction of monolayer iodinenes based on XB, with the usage of ingenious non-edge-to-edge tiling theory, to find all the structures as exhaustively as possible. Geometrically, it is evident that herringbone iodinene can be achieved by placing iodine atoms at each vertex of a hexagonal herringbone tiling, with the long edges representing the XBs. Therefore, we aim to construct iodinenes by bonding diatomic iodine molecules with XBs and utilizing the tiling theory~\cite{2Dcovalent2019tilings1,Tiling2014Phosphorene2,tessellated2017COF3}.

	The molecular orbital picture of the halogen bonding between two iodine molecules provides a valuable reference for constructing the halogen bonding network under investigation.  As shown in Fig.~\ref{fig1}(d), each molecule has one $\sigma$$_{p_x}$ bond, two $p_y$ orbitals (lone pairs) and one $\sigma^{*}_{p_x}$ antibond orbital~\cite{Hoffmann}. $s$ orbital forms a core pair and the $p_z$ orbital lies out of the plane, so they are ignored in the formation of XB. Electrons on the top diiodine could be transferred from $p_y$ lone pair to the $\sigma^{*}_{p_x}$ antibond orbital on the bottom diiodine, which is the orbital interaction picture of halogen bonding~\cite{XB2016review}. For the case of herringbone iodinene, every molecule acts as both an electron donor and acceptor, constituting an XB net (red dashed line in Fig.~\ref{fig1}(b)). Hence from the XB interaction picture in herringbone iodinene, several essential points can be derived to guide the construction of iodinene.  (1) Each diiodine molecule exhibits two $\sigma$-holes and two lone pairs within the same plane, enabling each atom to form two connections with other diiodines. Constructing monolayer iodinene based on the XB network is equivalent to a one-to-one correspondence between the $\sigma$-hole and the lone pair. (2) Iodinene is expected to be planar, as all $\sigma$-holes and lone pairs involved in the XB are confined to the same plane. (3) Every iodine atom should be considered equivalent to one another in principle. There is no justification for discrimination among them, as the XBs occur uniformly between diiodine molecules.

	The tiling theory can help us construct iodinenes systematically. If any two polygons are either disjoint or share one vertex or one entire edge in common, tiling by convex polygons is called edge-to-edge tiling~\cite{Grunbaum2Shephard,chavey1989tilings}. In contrast, if adjacent tiles share only part of their edge, this is so-called non-edge-to-edge tiling. Some works have predicted structures of 2D covalent materials utilizing the tiling theory~\cite{2Dcovalent2019tilings1,Tiling2014Phosphorene2,tessellated2017COF3}, more specifically, within the edge-to-edge classification. Regarding iodinenes, there are refined concepts that involve non-edge-to-edge tilings. Firstly, each atom is treated as a vertex, and intramolecular bonds and XBs are represented as edges in the tiling. Secondly, each vertex is connected to three edges, two of which have equal length, which corresponds to the case that each iodine atom connects a covalent bond and two equidistant XBs. Additionally, adjacent edges with equal lengths cannot be collinear, because $\sigma$-hole and $p_y$ lone pair on the same atom cannot be in alignment. Lastly, all vertices should be equivalent, meaning they are related by the symmetry of the tiling, known as vertex-transitive or uniform tiling~\cite{Grunbaum2Shephard,chavey1989tilings}. Based on this analysis, we can identify the required tilings from the existing non-edge-to-edge patterns~\cite{NINETY1ONE1ISOGONAL1TILINGS}. The results are presented in Fig.~\ref{fig2}.
	
	Five non-edge-to-edge tilings (see Fig.~\ref{fig2}(a1)$ \sim $(a5)) are selected from the existing 91 types of uniform tiling~\cite{NINETY1ONE1ISOGONAL1TILINGS} to map the tiling pattern to iodinene structure. Hence, the nomenclatures of those iodinenes could be labeled by their tiling patterns. The initial length of the short and long edges are set as 2.75 and 3.60 {\AA} respectively,  according to the intramolecular and XB length in the case of monolayer herringbone iodinene, and all XBs'angle (see sketch map at Fig.~\ref{fig1}(e)) are set as 180 $^{o}$. After structural relaxation, all the initial structures are slightly distorted. Bonding features could be seen from the charge density and electrostatic potential (ESP) map (Fig. S6 \& S7). The herringbone pattern (Fig.~\ref{fig2}(b1)) is the same as the foregoing herringbone iodinene. The Pythagorean tiling (Fig.~\ref{fig2}(a2)) is tessellated by two different size squares, whose name originates from the \emph{Pythagorean} theorem, so we call the corresponding structure as Pythagorean iodinene (Fig.~\ref{fig2}(b2)). Gyrated truncated hexagonal tiling (GTH, see Fig.~\ref{fig2}(a3)) is composed of regular hexagons and regular trigons, resembling the diatomic-kagome lattice topologically. As for the gyrated hexagonal tiling (GH, see (Fig.~\ref{fig2}(a4))), XB forms the entire edge of the hexagon and partial edge of the trigon, which is reversed from the case of GTH. The fifth pattern is truncated distorted trihexagonal tiling (TDT, see (Fig.~\ref{fig2}(a5))), which is composed of small regular trigons and big distorted truncated trigons. More details such as the symmetry and Wyckoff position of these iodinenes are shown in Table. S2. The phonon spectra are calculated (see Fig. S10), revealing that herringbone, Pythagorean, and GTH/diatomic-kagome iodinenes exhibit no imaginary frequency. GH displayed a small imaginary frequency near the $\Gamma$ point, but it could be eliminated with a mere 4$\%$ tensile strain. 
	
	To comprehensively explore the potential structures of monolayer iodinenes, we employed additional conventional methods to generate more 2D configurations. (a) We assume some simple lattices, including square, triangle, honeycomb, lieb, kagome, etc.(see Fig. S19). These iodinenes behave as conventional covalent 2D crystals where iodine serves as a polyvalent element and are highly energetic and dynamically unstable (see Fig. S21), part of which is even more energetic than gas diiodine that is not likely to form a stable freestanding crystal (see Table. S2). (b) We manually connect diiodine-based building blocks (see Fig. S8) solely through XBs, without employing the knowledge of the tiling theory. Nine configurations (see Fig. S9) are obtained, and two of them (``T4'' and ``T4+T4m'', as detailed in SM~\cite{SM}) didn't appear before. However, both ``T4'' and ``T4+T4m'' exhibit dynamic instability (Fig. S10). This result may be attributed to the violation of the requirement for all iodine atoms to be equivalent. (c) Additionally, a structure search using the particle swarm optimization (PSO) method~\cite{calyspo2012} is conducted (see Fig. S19). The PSO search identifies the emergence of herringbone and Pythagorean iodinenes in our design. However, several low-energy configurations, such as GTH and GH iodinenes, are not detected. The contrast highlights the distinct effectiveness of our design.
	
	Formation energy ($\Delta$E, meV/molecule) of all configurations derived is listed in Table. S2, with the energy of gas diiodine being the reference zero, where both M06-L+D3 and PBE functionals are used. Under the M06-L+D3 scheme, herringbone is the most low-energetic iodinene structure ($\Delta$E = -509), followed by Pythagorean ($\Delta$E = -457), ``T4+T4m'', TDT, GH, ``T4'' and GTH. These seven configurations consist of halogen-bonded iodine molecules and possess the lowest energy exactly. Conversely, high-energetic configurations are comprised of iodine molecules without XB nets, like ITSS (isosceles triangle snub square) and OR (octagon rhombus), or manifest as covalent crystals (refer to Fig. S20). These outcomes collectively demonstrate the genuine inclination of monolayer iodinenes toward arrangements formed through halogen-bonded iodine molecules, thereby substantiating the value and validity of our devised approach.

  \begin{figure}
	\includegraphics[width=8.6 cm]{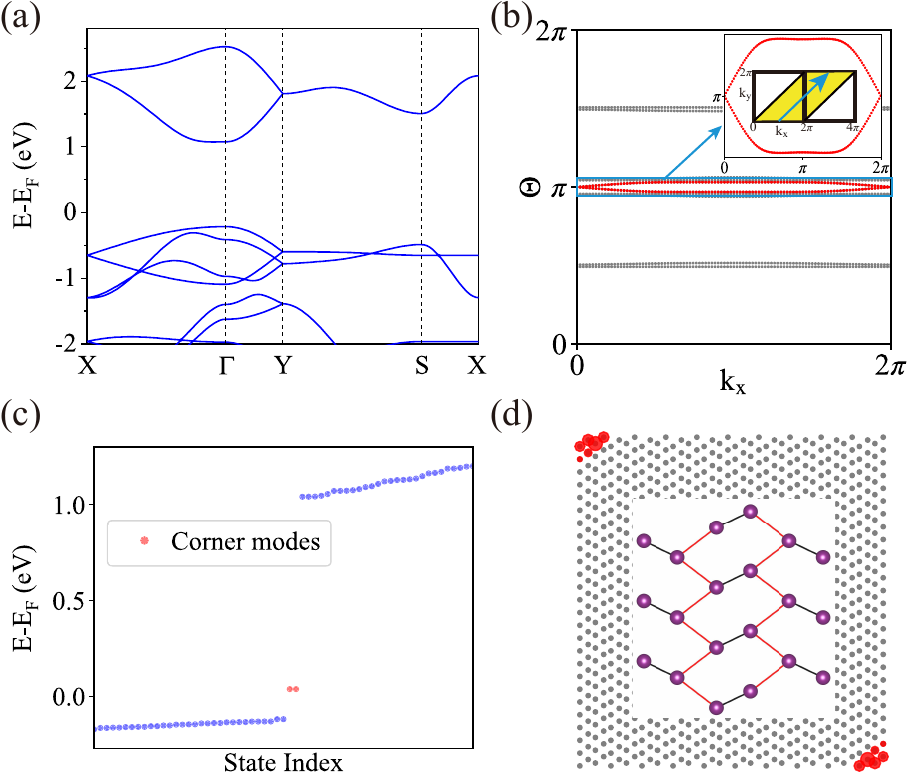}
	\caption{(a) Band structures of monolayer herringbone iodinene. (b) The Wilson loop along $k_{11}$ direction. Insets represent the enlarged Wilson loop around $\Theta=\pi$, indicating the nontrivial higher-order topology with the nonzero second Stiefel-Whitney number. (c) Energy spectra in real space, with the energy of the two corner states highlighted in red. (d) The distribution of the two corner states (depicted in red) in real space.}
	\label{fig3}
  \end{figure}

\textit{Band topology in monolayer iodinenes.--}
   We find herringbone, Pythagorean, GTH/diatomic-kagome and GH iodinenes are higher-order topological insulators. Due to both inversion symmetry and time-reversal symmetry are persevered in these four configurations, their higher-order topology can be characterized by the second Stiefel-Whitney number $w_2$~\cite{ahnBandTopologyLinking2018b}. Here, two methods are taken into consideration to prove their higher-order band topology: the parity criterion and the Wilson loop method. For the parity criterion, $w_2$ can be calculated by the parity eigenvalues of the valence bands, \begin{equation}
		(-1)^{w_2}=\Pi_{i=1}^4 (-1)^{\lfloor N_\text{occ}^- (\Gamma_i)/2\rfloor}\label{stiefel},
	\end{equation}
	where $\Gamma_i$ are the four time-reversal invariant momentums (TRIMs), $N_\text{occ}^-$ denotes the number of valence bands with odd parity and $\lfloor \cdots \rfloor$ is the floor function. Taking herringbone iodinene as an example, we find that $N_\text{occ}^-(\Gamma_i)$ are 6, 8, 7, and 7 for $\Gamma$, $M$, $X$ and $Y$, respectively, leading to a nontrivial second Stiefel-Whitney number $w_2=1$. This indicates that the herringbone iodinene is a   higher-order topological insulator. The nontrivial $w_2$ can also be evidenced by the Wilson loop method.
In Fig. \ref{fig3}(b), we plot the Wilson loop of the herringbone iodinene along  $k_{110}$ direction. One can find that the Wilson loop has only one crossing point on $\Theta=\pi$. This again proves the higher-order topology of the herringbone iodinene.
Therefore, the herringbone iodinene would have 0D-protected corner states. This is confirmed by our numerical calculation. The numerical result is plotted in Fig.~\ref{fig3}(c), from which two corner states in the band gap can be clearly observed. 
The charge density distributions of corner states in real space are presented in Fig.~\ref{fig3}(d). 
The other three kinds of iodinenes are also calculated as second Stiefel-Whitney insulators with nontrivial $w_2$ (see Fig. S16-18 and Table. S3).
	
Spin-orbit coupling (SOC) induced band topology is also studied considering the iodine element possesses a non-negligible SOC strength. As for the cases of Pythagorean, GTH/diatomic-kagome, and GH iodinenes, band crossings between the highest valence band and its lower band are all opened by SOC, which induces band inversions (see Fig.~\ref{fig4}). And the same behavior occurs between the lowest conduction band and its upper band. For spatial inversion invariant systems~\cite{topoSIS2007}, we can calculate the topological Z\textsubscript{2} invariant by counting the parity of the occupied states and find that these iodinenes are 2D topological insulators if the Fermi energy is shifted above the lowest unoccupied band or below highest occupied band. As shown in Fig.~\ref{fig4}(b), the gapless helical edge states connect the bulk states of GTH/diatomic-kagome iodinene, corresponding to the Z\textsubscript{2} values labeled in Fig.~\ref{fig4}(a), which originates from the bulk-boundary correspondence. Considering 2D materials are fabricated generally on substrates, the Fermi level could be shifted by choosing an appropriate substrate or doping. Hence these iodinenes could serve as a potential topological insulator.

	\begin{figure}
	\includegraphics[width=8.6 cm]{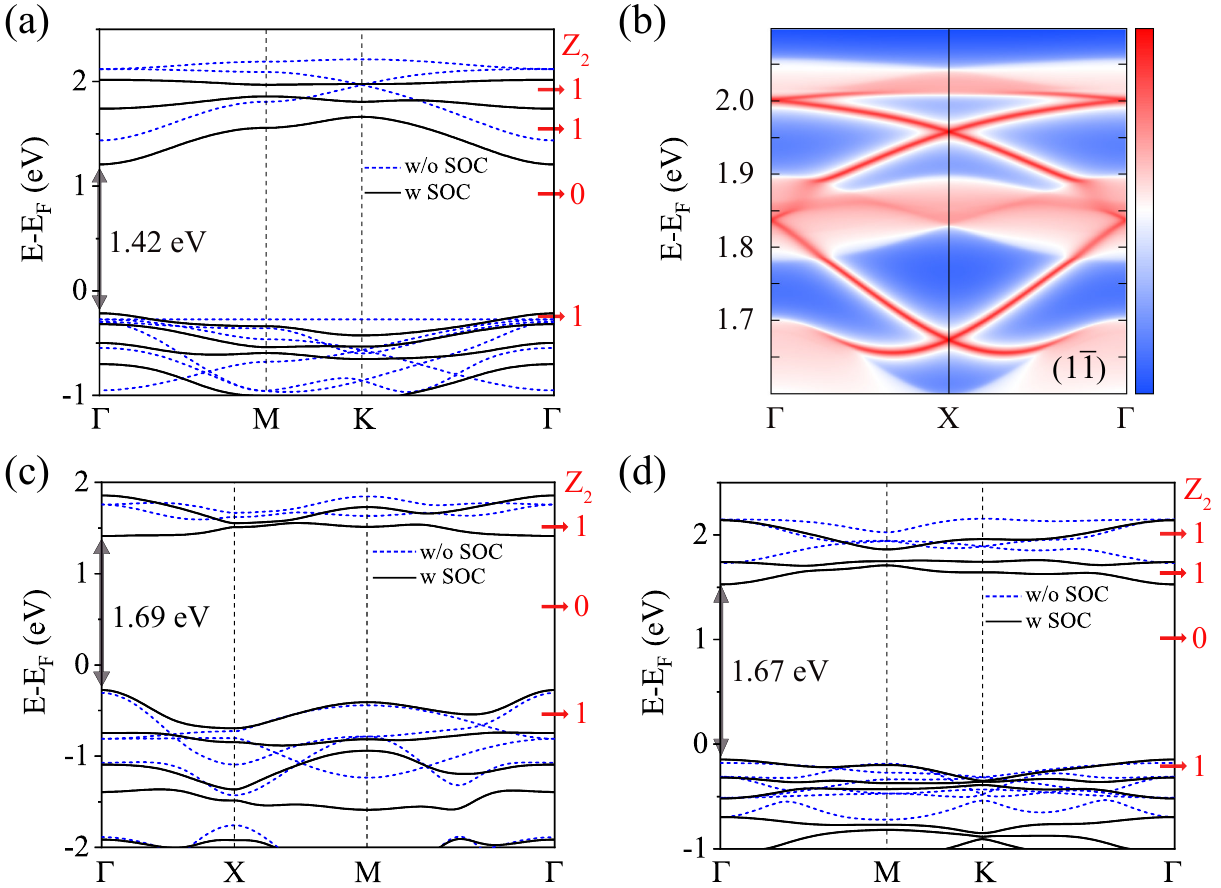}
	\caption{ (a) Band structures of GTH/diatomic-kagome iodinene. (b) Helical edge states of GTH/diatomic-kagome iodinene along the (1$\overline{1}$) edge. (c, d) Band structures of Pythagorean and GH iodinenes respectively. Blue dashed lines and black lines are the band structures without and with spin-orbit coupling (SOC) respectively. Grey arrows indicate the band gap. Topological Z\textsubscript{2} invariants are labeled above the occupied states defined. }
	\label{fig4}
	\end{figure}
\textit{Flat bands and optical absorption.--}
  As a noncovalent bond, XB could result in flat band structures in iodinenes because the relatively weak interactions between iodine molecules, and the flat band would bring about some novel properties. The band structures of herringbone iodinene are depicted in Fig. S11(a), where a quasi-direct band gap (1.165 eV) emerges between the conduction band minima (CBM) at $\Gamma$ and valence band maxima (VBM) near $\Gamma$. Notably, the highest occupied band from $\Gamma$ to VBM is almost flat with an energy shift of about only 8 meV (31 meV for HSE06), so it is a direct gap semiconductor approximately. The in-plane absorption coefficient curve is presented in Fig. S12(a), with two distinct peaks (around 2.04 and 2.74 eV) exceeding $3\times10^{5}$ $cm^{-1}$ in the visible region. This value is considerably higher than that observed in many other 2D materials and organic perovskite materials ($10^{4} \sim 10^{5} cm^{-1}$~\cite{huang2022ele_Optic}. As a result, herringbone iodinene is promising for the photoelectric conversion of solar energy.
  As shown in Fig.~\ref{fig4}, Pythagorean, GTH, and GH iodinenes are direct-gap semiconductors with the gap being 1.69, 1.42 and 1.67 eV respectively. Notably, GTH iodinene possesses a diatomic-kagome lattice leading to two flat bands~\cite{DiKagome1,DiKagome2} around -0.27 and 2.12 eV without SOC and one flat band around 2.0 eV with SOC. If with an appropriate substrate or heterojunction, the flat band is promising for a strong correlation platform. Besides, the computed in-plane optical absorption coefficients for Pythagoren, GTH, and GH iodinenes all exceed $ 10^{5}  cm^{-1} $  within the visible range (with peaks between $2\sim 3$ eV, see Fig. S12). These flat bands resulting from XBs and geometry contribute to a high density of states, and are beneficial to strong optical absorption, indicating their potential applications in photoelectronic devices.

\textit{Conclusion and discussion. ---}
	An innovative approach to the design of monolayer iodinenes has been put forward, involving the mapping of unique non-edge-to-edge tilings onto iodinene structures that consist of halogen-bonded iodine molecules. The effectiveness of this method has been validated through first-principles calculations, including the identification of four dynamically stable structures: herringbone, Pythagorean, GTH/diatomic-kagome, and GH. These iodinenes are confirmed as higher-order topological insulators with distinguishable corner states. And flat bands emerge, causing exceptional optical absorption within the visible light range. Furthermore, these iodinenes exhibit direct or quasi-direct band gaps, holding potential for optoelectronic application. Moreover, Pythagorean, GTH/diatomic-kagome, and GH iodinenes showcase nontrivial Z\textsubscript{2} band topology with appropriate doping. Notably, traditional approaches, including PSO searching, yield many structures devoid of halogen bonds. Those structures exhibit dynamic instability and higher energies than iodinenes formed by halogen-bonded iodine molecules. This outcome underscores the preference for iodinenes constructed from halogen-bonded iodine molecules, further substantiating the rationality of our design approach, and bringing about a new sight for the structural composition of 2D materials. 

	Different from other Xenes as covalent crystals, iodinene belongs to a new category of 2D crystals where XBs dominate. As XBs have extensive applications in crystal engineering~\cite{XB2020Crystal2dEngineering,XB2007Crystal2Engineering}, catalysis~\cite{XB2019Catalysis}, supramolecular chemistry~\cite{XB2013supramolecular2material,XB2001supramolecular2chemistry}, biomolecular systems~\cite{XB2004BiologicalMole}, self-assembly~\cite{XB2000Self2Assembly1,XB2016Self2Assembly2} and drug design~\cite{XB2009DrugDesign}, etc. already, monolayer iodinenes including herringbone, Pythagorean, GTH/diatomic-kagome and GH patterns would provide a new platform to explore more innovations.

\textit{Note added.} We become aware of an independent work recently~\cite{zhuIodene2023}. The work also studies the 2D sheet of iodine.	
	

%
\textit{Acknowledgements. ---}
	We would like to thank Jin Cao, Chaoxi Cui, Xiaodong Zhou, Xiuxian Yang, and Liping Liu for their helpful discussions. The work is supported by the National Key R\&D Program of China (Grant No. 2020YFA0308800) and the National Natural Science Foundation of China (Grant Nos. 12374055, 12204330). 
	
\bibliography{references}
%
\clearpage
\onecolumngrid
\begin{center}
	\textbf{\large Supplementary Material for\\ ``Design monolayer iodinenes based on halogen bond and tiling theory''}\\[.2cm]		
	Kejun Yu, $^{1,}$$^{2}$  Yichen Liu, $^{1,}$$^{2}$  Botao Fu, $^{3}$  Runwu Zhang, $^{1,}$$^{2}$  Da-shuai Ma, $^{4,}$$^{5}$  Xiao-ping Li, ${}^6$  Zhi-Ming Yu, $^{1,}$$^{2}$  Cheng-Cheng Liu, $^{1,}$$^{2}$  Yugui Yao, $^{1,}$$^{2}$\\[.1cm]
	{\itshape ${}^1$Centre for Quantum Physics, Key Laboratory of Advanced Optoelectronic Quantum Architecture and Measurement (MOE), School of Physics, Beijing Institute of Technology, Beijing, 100081, China}
	
	{\itshape ${}^2$Beijing Key Lab of Nanophotonics \& Ultrafine Optoelectronic Systems, School of Physics, Beijing Institute of Technology, Beijing, 100081, China}
	
	{\itshape ${}^3$College of Physics and Electronic Engineering, Center for Computational Sciences, Sichuan Normal University, Chengdu, 610068, China}
	
    {\itshape ${}^4$Institute for Structure and Function \& Department of Physics \& Chongqing Key Laboratory for Strongly Coupled Physics, Chongqing University, Chongqing 400044, P. R. China}
    
    {\itshape ${}^5$Center of Quantum materials and devices, Chongqing University, Chongqing 400044, P. R. China}
    
    {\itshape ${}^6$School of Physical Science and Technology, Inner Mongolia University, Hohhot 010021, China}
	
\end{center}

\maketitle

\setcounter{equation}{0}
\setcounter{section}{0}
\setcounter{figure}{0}
\setcounter{table}{0}
\setcounter{page}{1}
\renewcommand{\theequation}{S\arabic{equation}}
\renewcommand{\thesection}{ \Roman{section}}

\renewcommand{\thefigure}{S\arabic{figure}}
\renewcommand{\thetable}{\arabic{table}}
\renewcommand{\tablename}{Supplementary Table}

\renewcommand{\bibnumfmt}[1]{[S#1]}
\renewcommand{\citenumfont}[1]{#1}
\makeatletter

\maketitle

\setcounter{equation}{0}
\setcounter{section}{0}
\setcounter{figure}{0}
\setcounter{table}{0}
\setcounter{page}{1}
\renewcommand{\theequation}{S\arabic{equation}}
\renewcommand{\thesection}{ \Roman{section}}

\renewcommand{\thefigure}{S\arabic{figure}}
\renewcommand{\thetable}{\arabic{table}}
\renewcommand{\tablename}{Supplementary Table}

\renewcommand{\bibnumfmt}[1]{[S#1]}
\makeatletter

\maketitle
\subsection{Computational details}
\subsubsection{Parameters used in VASP and PSO}
Our electronic structure calculations are carried out using the Vienna ab initio simulation package (VASP)\cite{VASP1994} with the projector-augmented wave potential method\cite{PAWmethod1994}\cite{PAWmethod1999}. Plane-wave cutoff of 260 eV and Gamma-centered Monkhorst-Pack grid with a k-point resolution of more than 2$\pi$ $\cdot$ 0.04 {\AA}$^{-1}$ are utilized for the self-consistent calculation. Vacuum thickness is set as 20 {\AA} for 2D structural models. Structure parameters of the goundstate are obtained using the energy convergence criteria of $10^{-6}$ eV. For phonon spectra calculations, supercell size is chosen large enough, generally more than 3$\cdot$3$\cdot$1 of the unit cell. The finite displacement method is used for calculating the force constants and phonon dispersion was obtained by using Phonopy package\cite{phonopy2015}.  

\subsubsection{Choice of exchange-correlation functionals}
\begin{figure}[h]
	\centering
	\includegraphics[width=15cm]{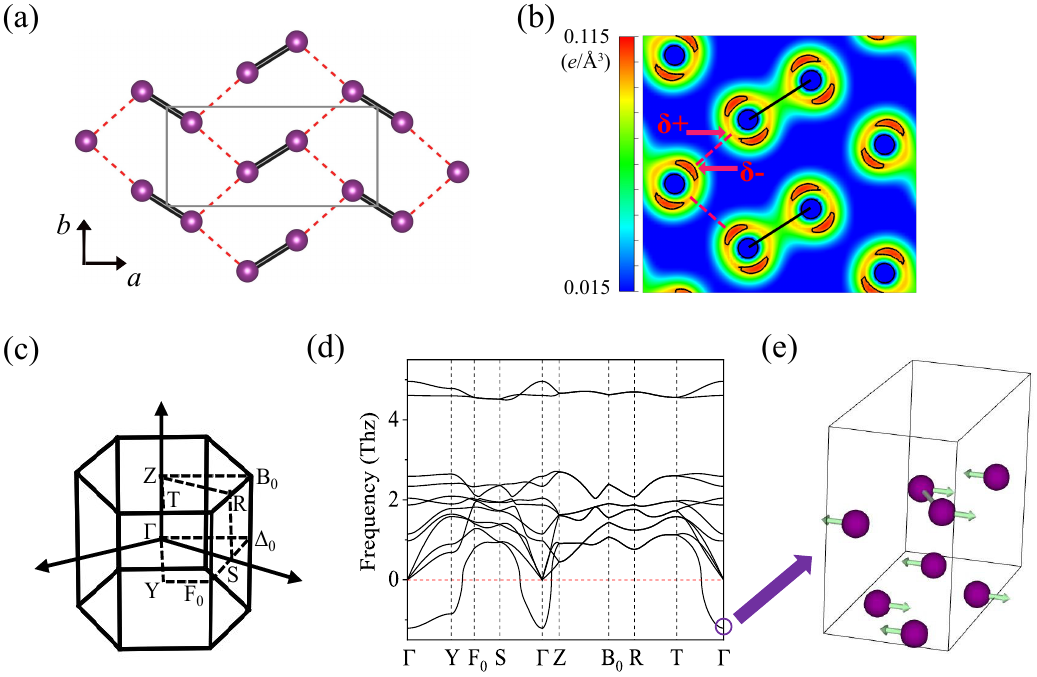}
	\caption{(a) Top view of bulk iodine. The length of the intramolecular bond (black lines) and the halogen bond (red dashed lines) is 2.76 and 3.55 Å, respectively. (b) Charge distribution in the top view of monolayers within bulk iodine. The charge density on the black crescent contour is 0.1 a.u., inside which lies a relatively higher charge density. ``$\sigma$-'' and ``$\sigma$+'' represent the electron concentration and depletion regions, respectively. Black lines indicate intramolecular covalent bonds, while red dot lines are for halogen bonds. (c) Brillouin zone for the primitive cell of bulk iodine. (d) Phonon spectra are calculated within the PBE+D3 scheme. (e) Vibration motion between iodine molecules along \emph{b} axis corresponding to the imaginary frequency of optical branch at $\Gamma$ in (d). }
	\label{figS1}
\end{figure}

As PBE functional is adopted widely in Density Functional Theory (DFT) calculations for solids, we test PBE\cite{PBE1996} functional first. Since bulk iodine is made up of diatomic iodine molecules, DFT-D3\cite{GrimmeDFTD3} was combined with PBE to include Van der Waals (VDW) interactions. Firstly, we calculated the phonon spectra of bulk iodine. An imaginary frequency of the optical branch appears at $\Gamma$ (see Fig.~\ref{figS1}(d)), which corresponds to slide motion between adjacent molecules along \emph{b} axis. As bulk iodine is stable actually, this motion reveals a missing attraction between adjacent molecules. Hence PBE is talentless for DFT calculations of bulk iodine, even combined with DFT-D3. This result is consistent with the fact that it's the halogen bond that dominates the interaction between adjacent molecules, not ordinary covalent bonds or VDW interactions. 

The next step is to find an appropriate functional for evaluating halogen bonds in bulk iodine. A DFT study of halogen bonding\cite{M062006L} indicates that the M06-L functional demonstrates capability in calculating periodic systems involving such interactions. As one of the Minnesota functionals, M06-L includes no exact exchange, which would be computational economically compared to hybrid functionals. M06-L has a good performance on medium-range noncovalent interactions (NCIs), which is likely to cover the halogen bond interactions. Moreover, considering the same rung on the Jacob's Ladder\cite{JaLadder2003PRL}, SCAN\cite{SCAN2015} is also tested. 

\begin{figure}[h]
	\centering
	\includegraphics[scale=0.68]{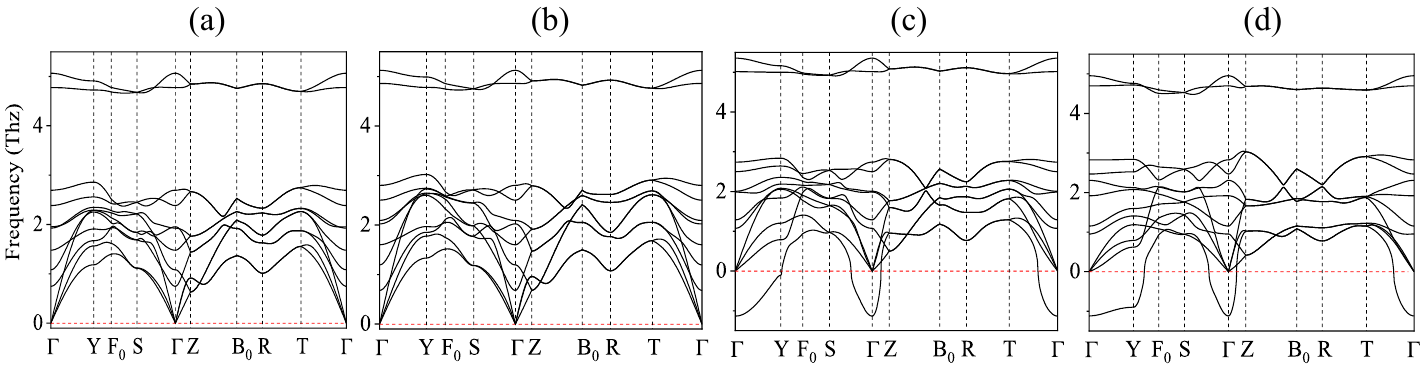}
	\caption{Phonon spectra of bulk iodine calculated with different functionals: a) M06-L, b) M06-L+D3, c) SCAN+D3, d) PBEsol.}
	\label{figS2}
\end{figure}

Lattice parameters relaxed within different DFT schemes are listed in Table.~\ref{tableS1}, from which we find that the best approximation has been achieved by M06-L+D3. Deviations from experiment (5K)\cite{SolidIod5K1992} are 2.1\%, 0.8\% and -2.3\% for \emph{a, b and c}, respectively. And deviation of conventional cell volume is 0.47\%. Fig.~\ref{figS2} shows the phonon spectra of bulk iodine calculated with M06-L, M06-L+D3, SCAN+D3, and PBEsol. Dynamic stability is achieved using M06-L or M06-L+D3, while PBEsol and SCAN present the same phonon imaginary frequency with PBE. We can conclude that M06-L could be a good choice for DFT calculations of halogen bonds at the meta-GGA level. And we choose M06-L+D3 for subsequent DFT calculations of 2D iodinenes.

\begin{table}[h]
	\caption{Lattice parameters (in unit Å) calculated within different DFT schemes.}
	\begin{tabular}{|c|c|c|c|c|c|c|c|c|}
		\hline 
		Lattice & PBE & PBE+D3 & M06-L & M06-L+D3 & SCAN & SCAN+D3 & PBEsol & \emph{Exp{*} 5K\cite{SolidIod5K1992}/110K\cite{SolidIod110K}}\tabularnewline
		\hline 
		\hline 
		a & 9.824 & 9.710 & 10.045 & 9.997 & 9.800 & 9.706 & 9.639 & \emph{9.796/9.805}\tabularnewline
		\hline 
		b & 4.607 & 4.559 & 4.750 & 4.698 & 4.576 & 4.526 & 4.380 & \emph{4.660/4.693}\tabularnewline
		\hline 
		c & 8.847 & 7.521 & 7.073 & 6.952 & 7.582 & 7.233 & 7.510 & \emph{7.119/7.160}\tabularnewline
		\hline 
	\end{tabular}
	\label{tableS1}
\end{table}

\subsubsection{Brillouin zones of 2D iodinenes}	
\begin{figure}[h]
	\centering
	\includegraphics[width=16cm]{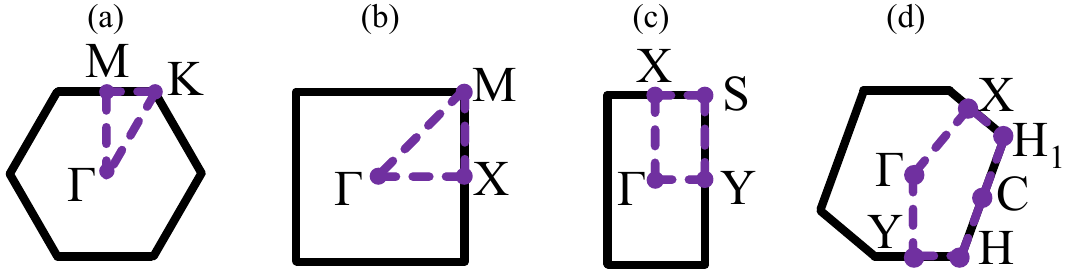}
	\caption{Schematic diagrams of 2D Brillouin zones of Bravais lattices: a) Hexagonal, b) Square, c) Rectangular, d) Oblique.}
	\label{figS10}
\end{figure}

As can be seen from Fig.~\ref{figS10}, Brillouin zones of various possible 2D configurations of iodinenes are shown.

\subsubsection{Numerical-noise issues in M06-L calculations}

\begin{figure}[htb]
	\centering
	\includegraphics[width=16cm]{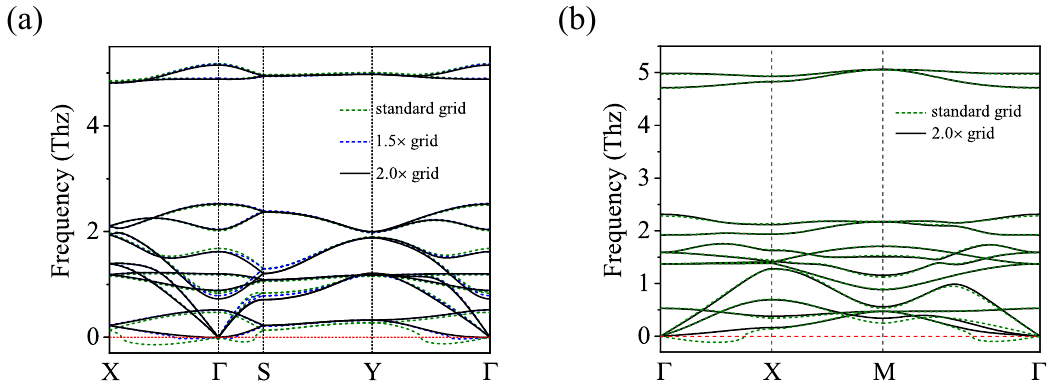}
	\caption{a) Phonon spectra of freestanding herringbone iodinene using different charge-density grids. b) Phonon spectra of freestanding Pythagorean iodinene using different charge-density grids.}
	\label{figS17}
\end{figure}

Phonon band structures calculated by Phonopy are derived from the force constants calculated by VASP. The second derivative of electron density serves as a variable for the exchange-correlation functional of meta-GGA. Small numerical noise in the electron density would be significantly amplified in the second derivative, so a more dense grid of electron density is needed to ensure accuracy while using M06-L, more specifically, increasing the NGX, NGY, and NGZ in VASP. As in the case of herringbone and Pythagorean iodinene (Fig.~\ref{figS17}), increasing the number of grid points along each dimension by 2$\times$ can remove the unphysical imaginary modes.

\subsubsection{Dynamic instability of herringbone and Pythagorean iodinene without the inclusion of halogen bond}
\begin{figure}[h]
	\centering
	\includegraphics[width=15cm]{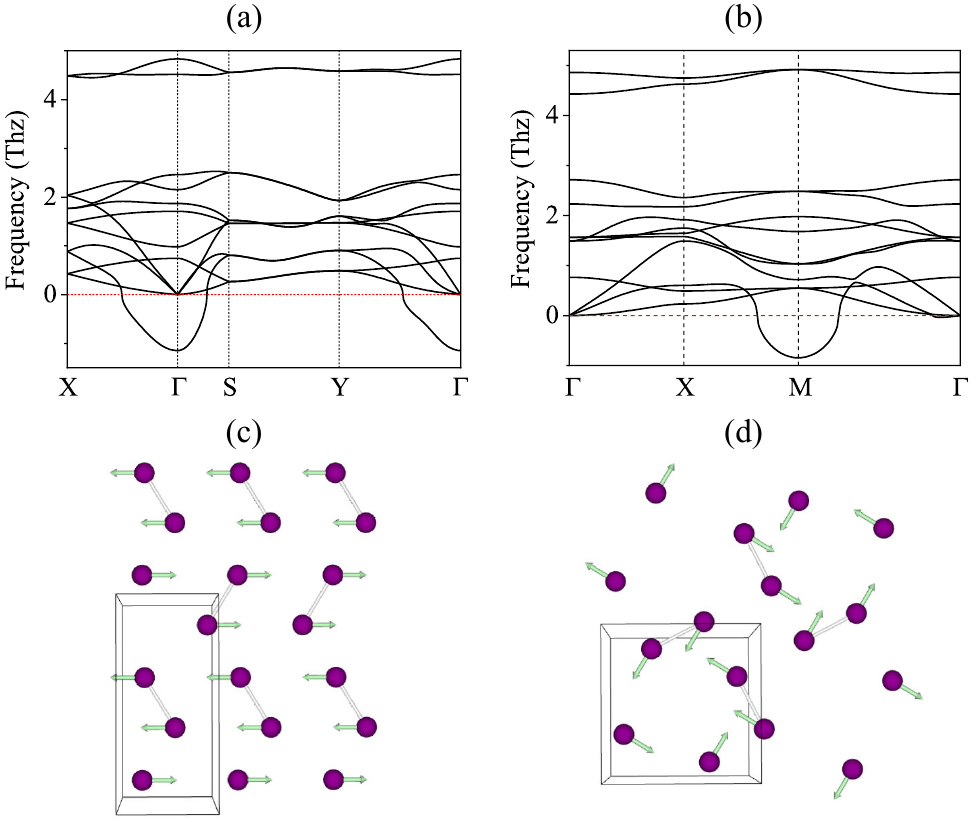}
	\caption{a), b) Phonon spectra of herringbone and Pythagorean iodinene calculated within PBE scheme, respectively. Imaginary frequency arises at $\Gamma$ for herringbone and M for Pythagorean iodinene. c) Eigen vibration mode of imaginary frequency at $\Gamma$ for herringbone iodinene. d) Eigen vibration mode of imaginary frequency at M for Pythagorean iodinene. }
	\label{figS12}
\end{figure}

Halogen bond is not able to be evaluated accurately in the PBE scheme when the 3D case of bulk iodine has been shown ahead. Herein we compare the phonon spectra within PBE and M06-L+D3 scheme for 2D crystals embodying halogen bond. As for 2D herringbone and Pythagorean iodinene, results are shown in Fig.~\ref{figS12} and Fig.~\ref{figS17}, respectively. The comparison shows that halogen bonds are essential for the stability of crystals of iodine, no matter 3D or 2D cases.

\subsection{Construct monolayer iodinenes from non edge-to-edge tilings and building blocks}

\begin{figure}[h]
	\centering
	\includegraphics[scale=0.75]{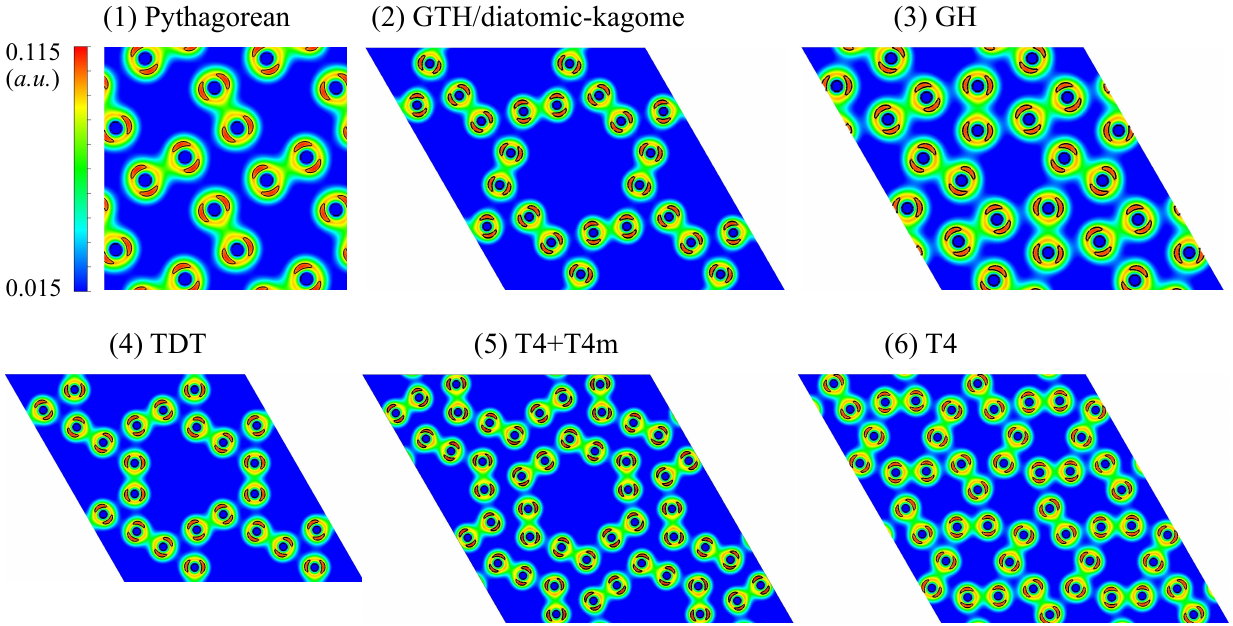}
	\caption{Valence electron charge density of iodinenes. (1-4) The valence electron charge density of iodinenes constructed from non edge-to-edge tilings: Pythagorean, GTH, GH, TDT. (5-6) The
		valence electron charge density of iodinenes constructed from building blocks: ``T4+T4m'', ``T4''. The unit of charge density is an atomic unit (e/bohr\textasciicircum{}3), and density inside the black crescent contour is > 0.1 a.u..}
	\label{figS6}
\end{figure}

\begin{figure}[h]
	\centering
	\includegraphics[scale=0.75]{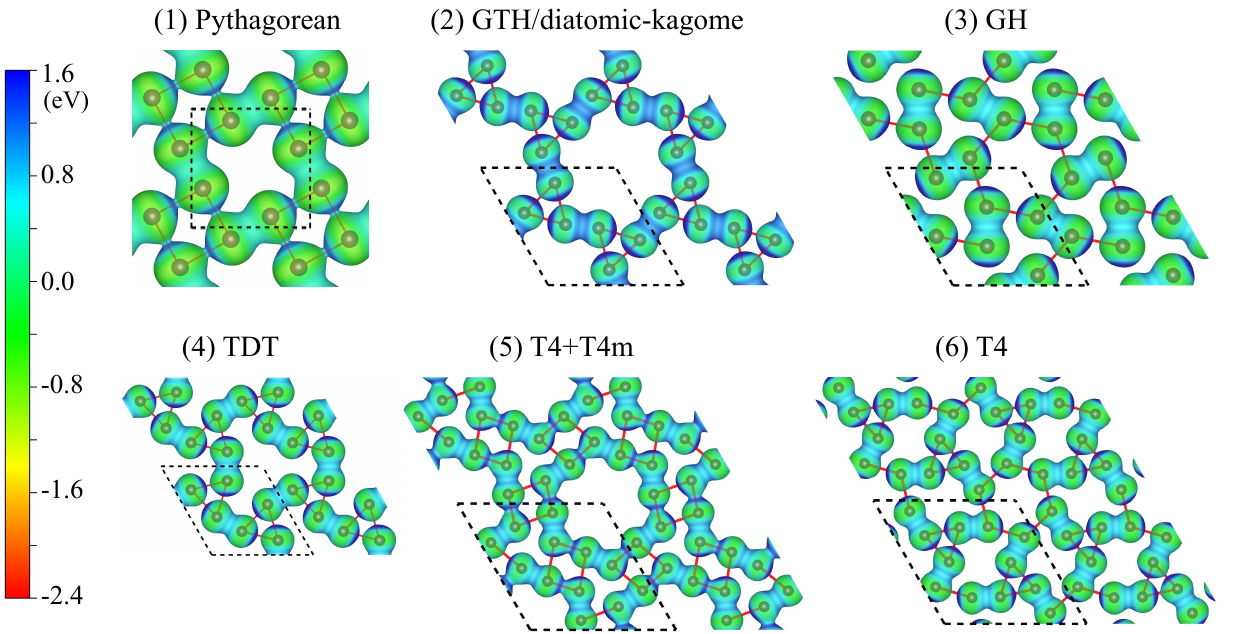}
	\caption{Electrostatic potential (ESP) map of iodinenes. The ESP on 0.015 a.u. isosurface of charge density. (1-4) correspond to iodinenes mapped from non edge-to-edge tilings: Pythagorean, GTH, GH, TDT. (5-6) correspond to iodinenes constructed from building blocks: ``T4+T4m'', ``T4''. }
	\label{figS7}
\end{figure}

Firstly, we show the geometrical orientation of XBs in iodinenes. Figures~\ref{figS6}(1-4) are the valence electron charge density of Pythagorean, GTH, GH, and TDT iodinenes, respectively. Electrostatic potential (ESP) maps of all iodinenes in Fig.~\ref{figS7} reveal that the electron-depleted areas are truly accompanied by positive electrostatic potential, i.e., the $\sigma$-hole. An XB occurs between $\sigma$-hole and an adjacent lone pair which is surrounded by negative electrostatic potential. Then we can see diiodines are hinged with each other by XBs (the red lines in Fig.~\ref{figS7}).


\begin{figure}[h]
	\centering
	\includegraphics[scale=0.8]{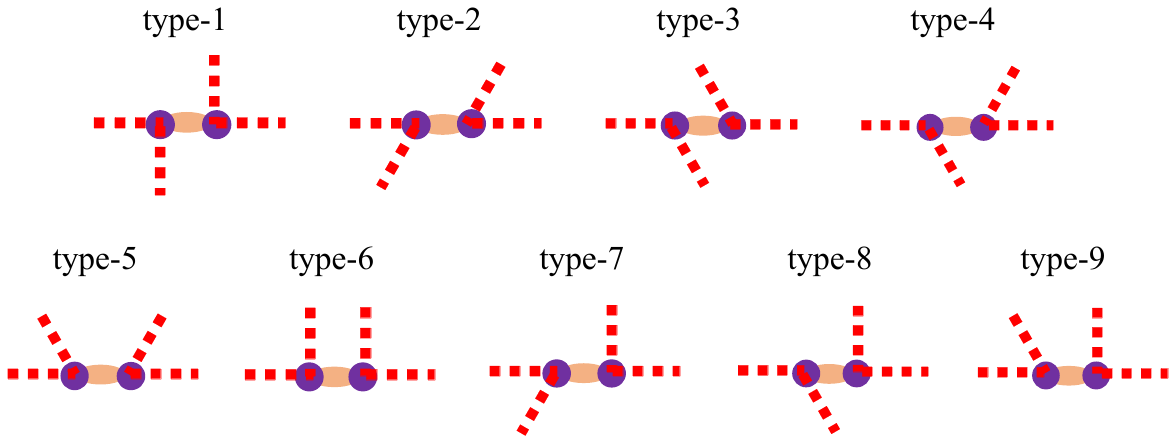}
	\caption{Basic building blocks. Type-1$ \sim $9 are building blocks made from diiodine, while potential XBs with rational orientation are represented by the red dot lines.}
	\label{figS8}
\end{figure}

\begin{figure}[h]
	\centering
	\includegraphics[scale=0.7]{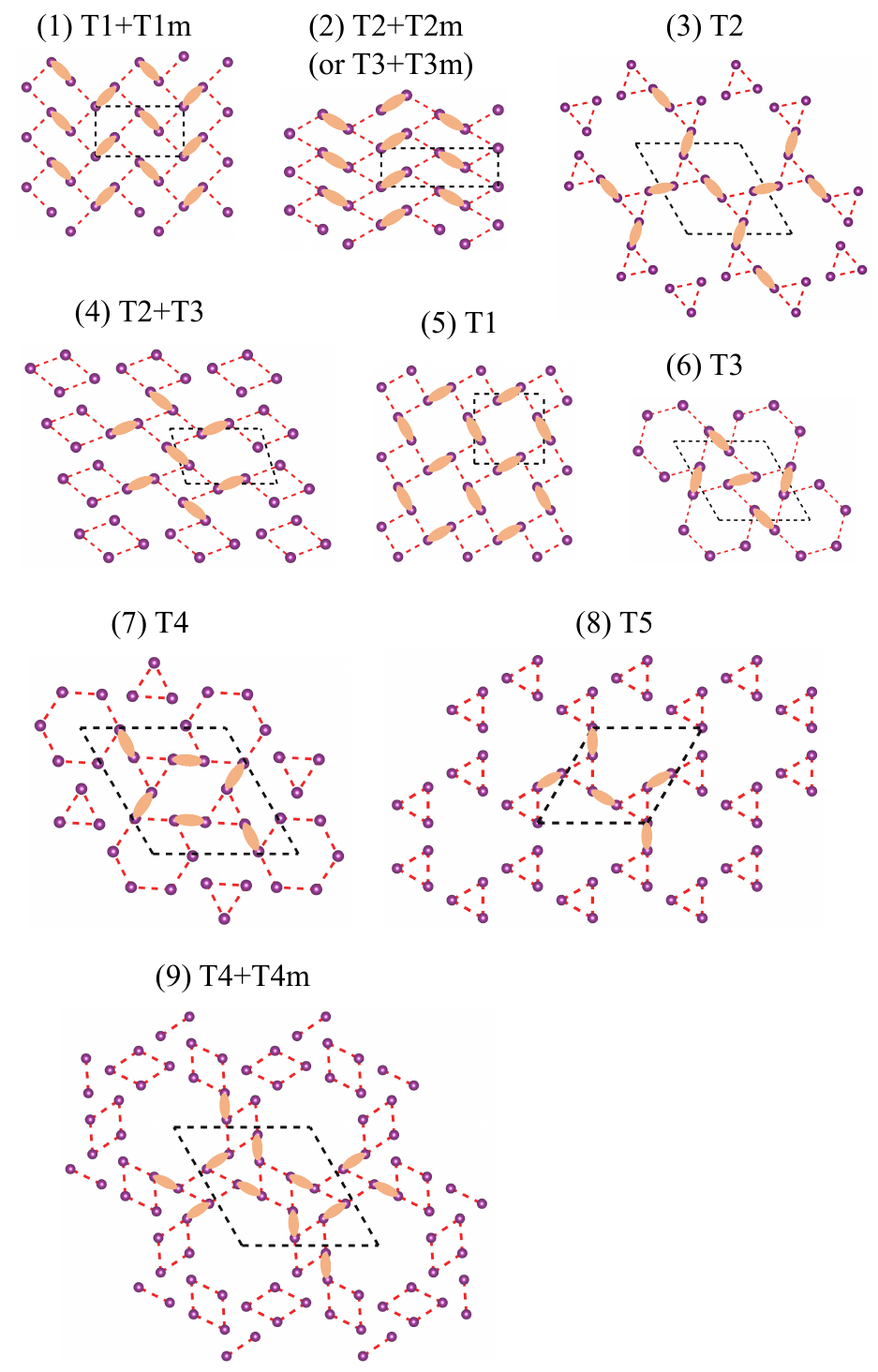}
	\caption{Iodinenes spliced from basic building blocks. ``Ti'' is spliced from ``type-i'' (i=1, 2, 3, 4, 5) building block (as depicted in Fig.~\ref{figS8}). ``Ti+Tjm'' means that the configuration is composed of ``type-i'' and ``type-jm'', while ``type-jm'' is the mirror of ``type-j''.}
	\label{figS9}
\end{figure}

A comprehensive combination of basic building blocks is done to avoid missing some potential tiling patterns when distortion is allowed. We construct monolayer iodinenes from building blocks step by step: Firstly, basic building blocks need to be determined. As every atom has two hands to bond with other diiodines and to consider periodicity, the initial angle between XB on the same atom should be set to 60$^\circ$, 90$^\circ$, 120$^\circ$. There are 9 basic building blocks (see Fig.~\ref{figS8}), but to form a periodic structure with the rational arrangement of diiodines, type-6, type-7, type-8 and type-9 blocks need to be rejected. The remaining building blocks could be used to construct iodinenes, whose framework is shown in Fig.~\ref{figS9} visually. In Fig.~\ref{figS8} and Fig.~\ref{figS9}, purple balls, yellow ellipses, and red dot lines represent the iodine atoms, covalent bonds, and XBs respectively. We just nominate those configurations by the building blocks inside themselves. For example, ``T1'' (see Fig.~\ref{figS9}(5)) is spliced from ``type-1'' building block. ``T1+T1m'' (see Fig.~\ref{figS9}(1)) is spliced from ``type-1'' and the mirror of ``type-1'', while ``m'' represents mirror. 

All tilings aforementioned could be rebuilt from those building blocks. It could be seen from Fig.~\ref{figS9} that: (1, 5, 3, 6, 8) are the rectangle herringbone, Pythagorean, GTH, GH, TDT tiling pattern. Moreover, some new structures arise (Fig.~\ref{figS9}(2, 4, 7, 9)). After relaxation in DFT calculations, rectangle and parallelogram herringbone (Fig.~\ref{figS9}(1)\&(2)) would evolve to hexagonal herringbone (the configuration of herringbone iodinene), which are topologically equivalent to each other. The same story could be told for rhombus Pythagorean (Fig.~\ref{figS9}(4)) and Pythagorean structure (Fig.~\ref{figS9}(5)). As shown in Fig.~\ref{figS9}, ``T4''(Fig.~\ref{figS9}(7)) is made from type-4 building block. ``T4+T4m''(Fig.~\ref{figS9}(9)) is made from type-4 building block and its mirror. For iodinenes that could be mapped from ununiform non edge-to-edge tilings, i.e. not all atoms are equivalent under symmetry operation, such as ``T4'' and ``T4+T4m'' (Fig.~\ref{figS9}(7 \& 9)), they bear fair-sized XB distortion during relaxation and are dynamically unstable revealed on phonon spectra (Fig.~\ref{figS11}(5 \& 6)).

\subsection{Stability of 2D monolayer iodinenes}

\begin{figure}[h]
	\centering
	\includegraphics[scale=1.0]{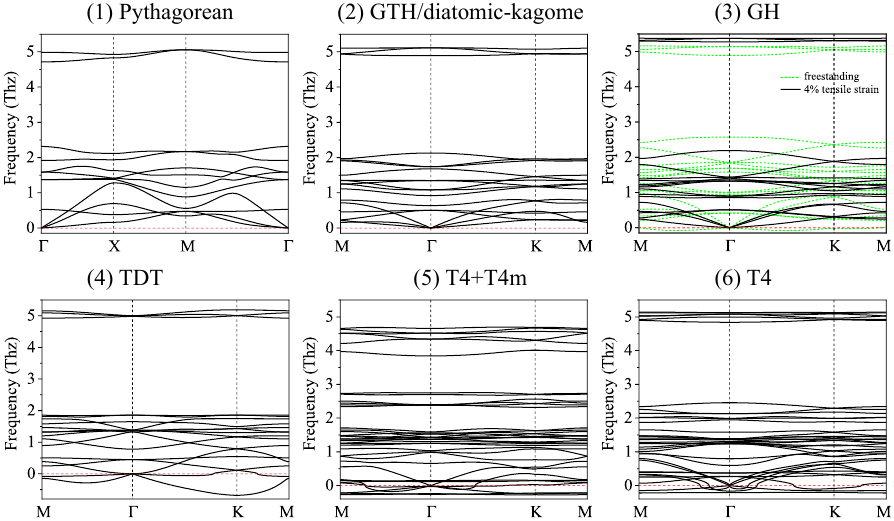}
	\caption{Phonon spectra of monolayer iodinenes designed from tilings and building blocks. (1-4): The phonon spectra of Pythagorean, Gyrated truncated hexagonal (GTH), Gyrated hexagonal (GH), and Truncated distorted trihexagonal (TDT) iodinenes. In (3), imaginary frequency arises for freestanding (green dash curves) GH iodinene but disappears when 4\% tensile strain
		(black solid curves) is applied. (5-6): The phonon spectra of ``T4+T4m'' and ``T4'' iodinenes.}
	\label{figS11}
\end{figure}

We show the crystal parameters and formation energies (Table.~\ref{tableS2}) of monolayer iodinenes. Phonon spectra of iodinenes constructed from diiodines are shown in Fig.~\ref{figS11}, while Fig.~\ref{figS16} from the PSO searching and simple lattice. In Table.~\ref{tableS2}, all structures including bulk iodine are listed in the order from the lowest energy to the highest energy. Both M06-L+D3 and PBE functionals are used to evaluate the influence of XB. As a wrong result to in turn highlight the importance of halogen bond, it is noteworthy that Pythagorean iodinene is less energetic than bulk iodine and herringbone iodinene in the PBE scheme, while similar results are present in the work of Ahmad et al. ~\cite{Pytha1iodinene2021}, where PBEsol was used and herringbone iodinene was demonstrated as an unstable structure. 

\begin{table}
	\begin{centering}
		\caption{Formation energies and crystal parameters of iodinenes. Defination of formation energy: $\Delta$E = $\frac{1}{n}$E -E$_m$ (meV). n = N/2, N is the number of iodine atoms in a unit cell; E$_m$ is the free energy of an isolated iodine molecule, serving as a reference level. (a, b, $\gamma$) are the length of primitive vectors and the angle between them. The number in parentheses in the fourth column illustrates the formation energy order of the corresponding iodinene based on the PBE scheme, arranged from the lowest to the highest value.} 
		\label{tableS2}    
		\par\end{centering}
	\resizebox{\linewidth}{!}{	
		\begin{tabular}{|c|c|c|c|c|c|c|c|}
			\hline 
			No. & Structures & $-\Delta$E/meV & $-\Delta$E/meV & space group & a, b, $\gamma$  & Wyckoff position  & a, b, $\gamma$ \tabularnewline
			\hline 
			&  & M06-L+D3 & PBE &  & M06-L+D3 & M06-L+D3 & PBE\tabularnewline
			\hline 
			\hline 
			0 & Bulk & 1188 & 380 & Cmce (64) &  &  & \tabularnewline
			\hline 
			1 & Herringbone & 509 & 369 (5) & Pbam (55) & 4.38, 10.66, 90$^\circ$ & 4h(0.137, 0.116, 0.5) & 4.59, 9.84, 90$^\circ$\tabularnewline
			\hline 
			2 & Pythagorean & 457 & 412 (2) & P4/m (83) & 7.18, 7.18, 90$^\circ$ & 4k(0.096, 0.323, 0.5) & 7.06, 7.06, 90$^\circ$\tabularnewline
			\hline 
			3 & T4+T4m & 422 & 420 (1) & P2/m (10) & 14.16, 14.16, 120$^\circ$ & 2n(0.184, 0.487, 0.5) & 14.15, 14.15, 120$^\circ$\tabularnewline
			&  &  &  &  &  & 2n(0.513, 0.697, 0.5) & \tabularnewline
			&  &  &  &  &  & 2n(0.690, 0.615, 0.5) & \tabularnewline
			&  &  &  &  &  & 2n(0.075, 0.690, 0.5) & \tabularnewline
			&  &  &  &  &  & 2n(0.303, 0.816, 0.5) & \tabularnewline
			&  &  &  &  &  & 2n(0.615, 0.925, 0.5) & \tabularnewline
			\hline 
			4 & TDT & 407 & 391 (3) & P-62m (189) & 9.7, 9.7, 120$^\circ$ & 6k(0.265, 0.429, 0.5) & 10.09, 10.09, 120$^\circ$\tabularnewline
			\hline 
			5 & GH & 391 & 292 (10) & P6/m (175) & 9.51, 9.51, 120$^\circ$ & 6k(0.419, 0.333, 0.5) & 9.17, 9.17, 120$^\circ$\tabularnewline
			\hline 
			6 & T4 & 383 & 347 (6) & P2/m (10) & 13.81, 13.81, 120$^\circ$ & 2n(0.087, 0.300, 0.5) & 13.41, 13.41, 120$^\circ$\tabularnewline
			&  &  &  &  &  & 2n(0.300, 0.212, 0.5) & \tabularnewline
			&  &  &  &  &  & 2n(0.502, 0.218, 0.5) & \tabularnewline
			&  &  &  &  &  & 2n(0.782, 0.283, 0.5) & \tabularnewline
			&  &  &  &  &  & 2n(0.717, 0.498, 0.5) & \tabularnewline
			&  &  &  &  &  & 2n(0.212, 0.913, 0.5) & \tabularnewline
			\hline 
			7 & GTH & 358 & 389 (4) & P6/m (175) & 11.21, 11.21, 120$^\circ$ & 6k(0.130, 0.614, 0.5) & 10.96, 10.96, 120$^\circ$\tabularnewline
			\hline 
			8 & Rectangle & 344 & 223 (13) & Pmmm (47) & 3.04, 3.78, 90$^\circ$ & 1c(0.0, 0.0, 0.5) & 3.01, 3.64, 90$^\circ$\tabularnewline
			\hline 
			9 & ITSS & 340 & 130 (16) & P4/mbm (127) & 7.14, 7.14, 90$^\circ$ & 4h(0.865, 0.365, 0.5) & 7.08, 7.08, 90$^\circ$\tabularnewline
			\hline 
			10 & $\alpha$-1 & 297 & 314 (9) & P2/m (10) & 6.22, 9.76, 100.33$^\circ$ & 2n(0.867, 0.360, 0.5) & 6.71, 10.78, 95.44$^\circ$\tabularnewline
			\hline 
			11 & $\alpha$-2 & 263 & 316 (8) & Cmmm (65) & 6.65, 6.65, 120$^\circ$ & 2i(0.257, 0.743, 0.5) & 6.73, 6.73, 120$^\circ$\tabularnewline
			&  &  &  &  &  & 1a(0.0, 0.0, 0.5) & \tabularnewline
			\hline 
			12 & $\alpha$-3 & 234 & 253 (12) & Pm (6) & 6.66, 8.81, 90$^\circ$ & 1b(0.441, 0.143, 0.5) & 6.43, 8.99, 91.28$^\circ$\tabularnewline
			&  &  &  &  &  & 1b(0.086, 0.803, 0.5) & \tabularnewline
			&  &  &  &  &  & 1b(0.249, 0.449, 0.5) & \tabularnewline
			&  &  &  &  &  & 1b(0.603, 0.789, 0.5) & \tabularnewline
			&  &  &  &  &  & 1b(0.845, 0.296, 0.5) & \tabularnewline
			\hline 
			13 & $\alpha$-4 & 226 & 331 (7) & Pm (6) & 6.62, 11.51, 106.6$^\circ$ & 1b(0.362, 0.848, 0.5) & 6.48, 11.57, 73.6$^\circ$\tabularnewline
			&  &  &  &  &  & 1b(0.004, 0.621, 0.5) & \tabularnewline
			&  &  &  &  &  & 1b(0.707, 0.372, 0.5) & \tabularnewline
			&  &  &  &  &  & 1b(0.231, 0.075, 0.5) & \tabularnewline
			&  &  &  &  &  & 1b(0.184, 0.325, 0.5) & \tabularnewline
			\hline 
			14 & $\alpha$-5 & 176 & 274 (11) & Amm2 (38) & 8.3, 8.3, 83.06$^\circ$ & 2e(0.664, 0.164, 0.5) & 8.97, 8.97, 95.97$^\circ$\tabularnewline
			&  &  &  &  &  & 2e(0.932, 0.396, 0.5) & \tabularnewline
			&  &  &  &  &  & 3b(0.664, 0.664, 0.5) & \tabularnewline
			\hline 
			15 & Kagome & 158 & 85 (17) & P6/mmm (191) & 6.57, 6.57, 120$^\circ$ & 3g(0.5, 0.0, 0.5) & 6.44, 6.44, 120$^\circ$\tabularnewline
			\hline 
			16 & Square & 126 & 160 (15) & P4/mmm (123) & 3.32, 3.32, 90$^\circ$ & 1b(0.0, 0.0, 0.5) & 3.24, 3.24, 90$^\circ$\tabularnewline
			\hline 
			17 & OR & 58 & -127 (18) & P4/mmm (123) & 8.58, 8.58, 90$^\circ$ & 4o(0.5, 0.156, 0.5) & 7.77, 7.77, 90$^\circ$\tabularnewline
			\hline 
			18 & Lieb & 43 & 166 (14) & P4/mmm (123) & 6.18, 6.18, 90$^\circ$ & 1b(0.0, 0.0, 0.5) & 6.07, 6.07, 90$^\circ$\tabularnewline
			&  &  &  &  &  & 2e(0.5, 0.0, 0.5) & \tabularnewline
			\hline 
			19 & Triangle & -144 & -194 (19) & P6/mmm (191) & 3.49, 3.49, 120$^\circ$ & 1b(0.0, 0.0, 0.5) & 3.37, 3.37, 120$^\circ$\tabularnewline
			\hline 
			20 & Honeycomb & -358 & -223 (20) & P6/mmm (191) & 5.67, 5.67, 120$^\circ$ & 2d(1/3, 2/3, 0.5) & 5.55, 5.55, 120$^\circ$\tabularnewline
			\hline 
			
		\end{tabular}
	}
\end{table}

\subsection{Electronic structure of herringbone, Pythagorean, GTH/diatomic-kagome and GH iodinenes}		

\subsubsection{Higher-order topological states in Pythagorean, GTH/diatomic-kagome and GH iodinenes}
\begin{figure}[h]
	\centering
	\includegraphics[width=11cm]{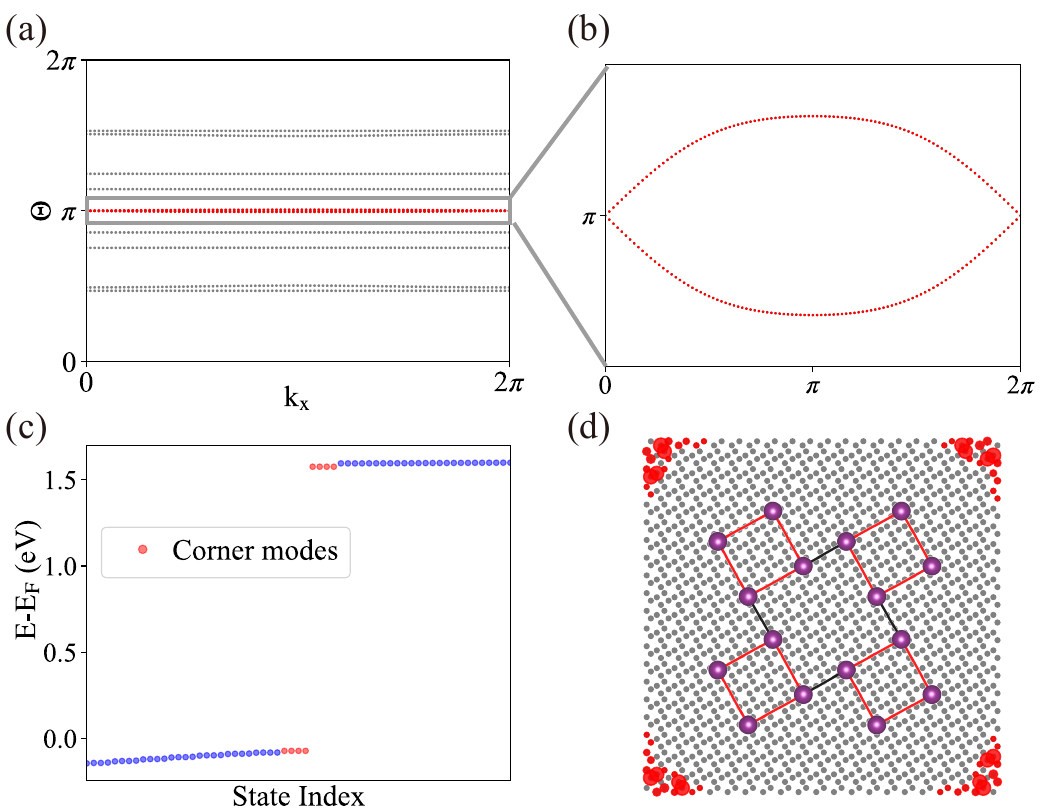}
	\caption{(a) Wilson loop of Pythagorean iodinene along $k_{110}$ and (b) a magnified view of its behavior around $\Theta=\pi$. (c) shows the energy spectra in real space and the corner states are highlighted in red. (d) shows the distribution of the corner states in real space. }
	\label{pytha}
\end{figure}

\begin{figure}[h]
	\centering
	\includegraphics[width=11cm]{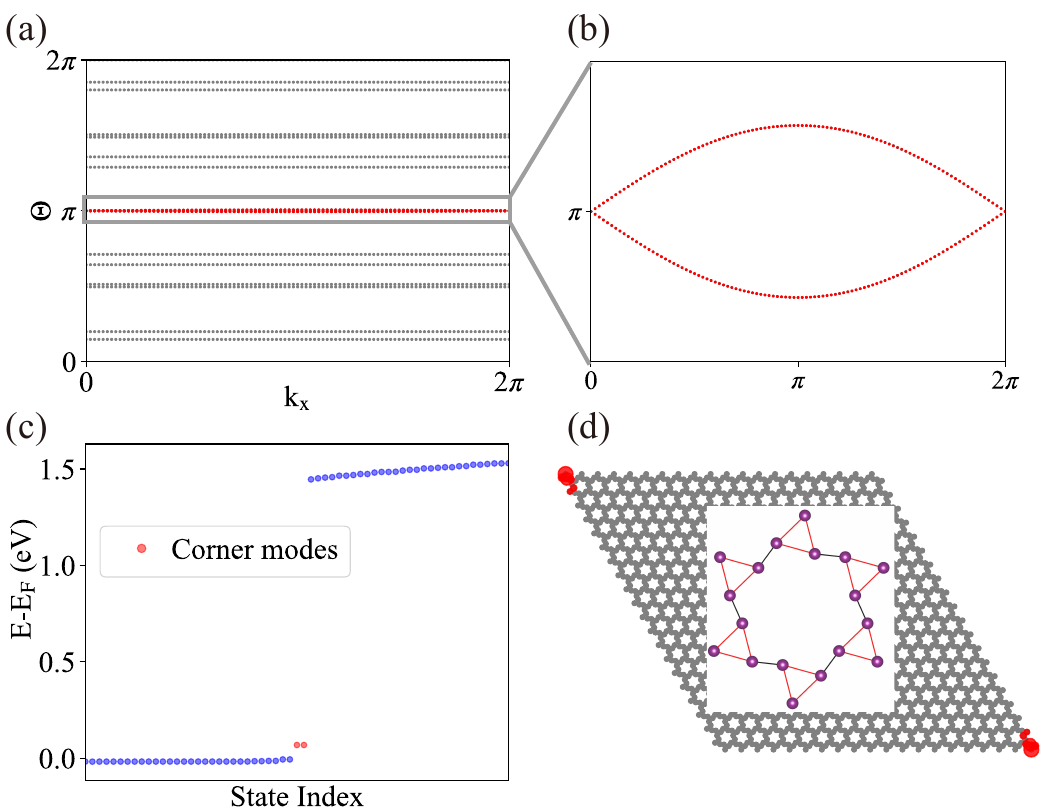}
	\caption{(a) Wilson loop of GTH/diatomic-kagome iodinene along $k_{110}$ and (b) a magnified view of its behavior around $\Theta=\pi$. (c) shows the energy spectra in real space and the corner states are highlighted in red. (d) shows the distribution of the corner states in real space. }
	\label{GTH-diKagome}
\end{figure}

\begin{figure}[h]
	\centering
	\includegraphics[width=11cm]{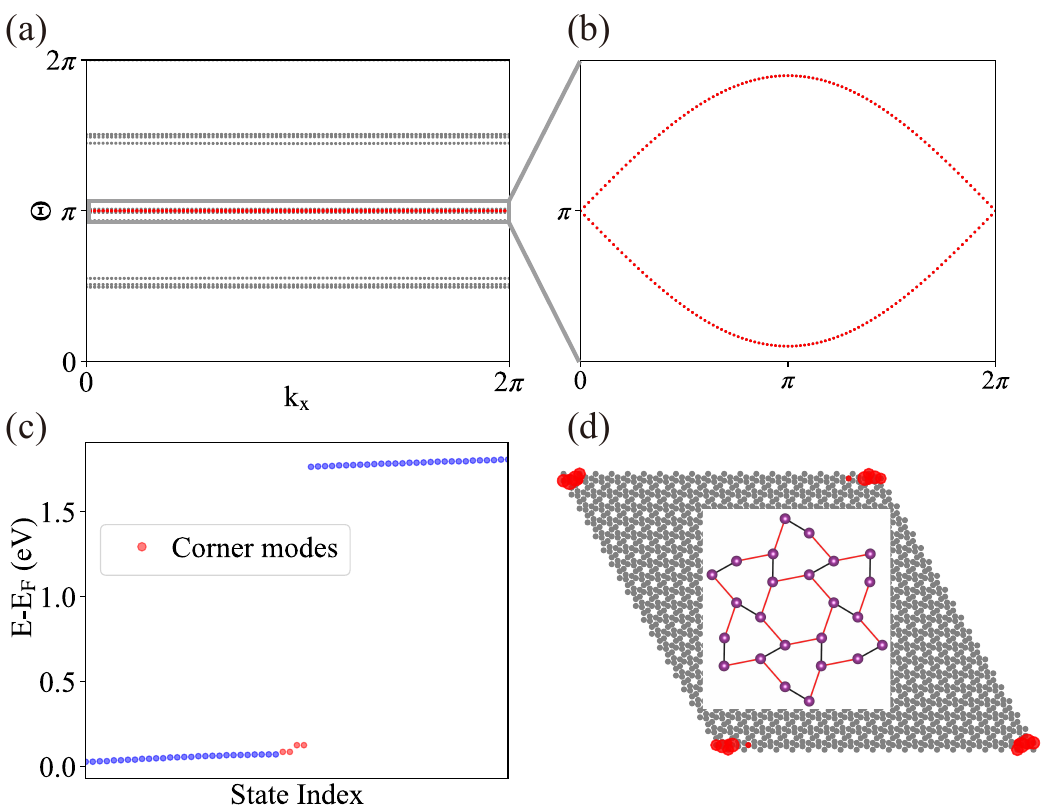}
	\caption{(a) Wilson loop of GH iodinene along $k_{110}$ and (b) a magnified view of its behavior around $\Theta=\pi$. (c) shows the energy spectra in real space and the corner states are highlighted in red. (d) shows the distribution of the corner states in real space. }
	\label{GH-t3}
\end{figure}

The Wilson loop evidence for the higher-order topological states of Pythagorean, GTH/diatomic-kagome, and GH iodinenes, as well as their energy spectra and real-space corner states, are presented in Fig.~\ref{pytha}, Fig.~\ref{GTH-diKagome} and Fig.~\ref{GH-t3}, respectively. The Wilson loop analysis indicates that all these iodinenes possess non-trivial topological invariant with Stiefel-Whiteny number $w_2=1$. As shown in Table.~\ref{tableS3}, we also get consistent results with the parity criterion. By evaluating the energy spectra of finite-size samples in real space, we further confirm that all iodinene structures exhibit topologically protected corner states.

\begin{table}[h]
	\begin{centering}
		\caption{Number of occupied bands with odd parity at TRIM for herringbone, Pythagorean, GTH/diatomic-kagome, and GH iodinenes. And $w_2$ is the Stiefel-Whiteny number.} 
		\label{tableS3}    
		\par\end{centering}
	\begin{tabular}{|c|c|c|c|c|c|}
		\hline 
		Structures & $\Gamma$ &$M$ & $X$ & $Y$& $w_2$ \\ 
		\hline 
		herringbone & 6 & 8 & 7 & 7 & 1\\
		\hline 
		Pythagorean & 6 & 8 & 7 & 7 & 1\\
		\hline 
		GTH/diatomic-kagome & 9 & 11 & 11 & 11 & 1\\
		\hline 
		GH & 9 & 11 & 11 & 11 & 1\\
		\hline 
	\end{tabular}
\end{table}

\subsubsection{$Z_2$ topological state of Pythagorean, GTH/diatomic-kagome and GH iodinenes}
WANNIER90 code\cite{wan90} is employed to get model Hamiltonian constructed from maximally localized Wannier functions (MLWFs),  and surface state calculations are implemented in the WannierTools package\cite{wanniertools} based on the Green functions method.

\begin{figure}[h]
	\centering
	\includegraphics[scale=1.0]{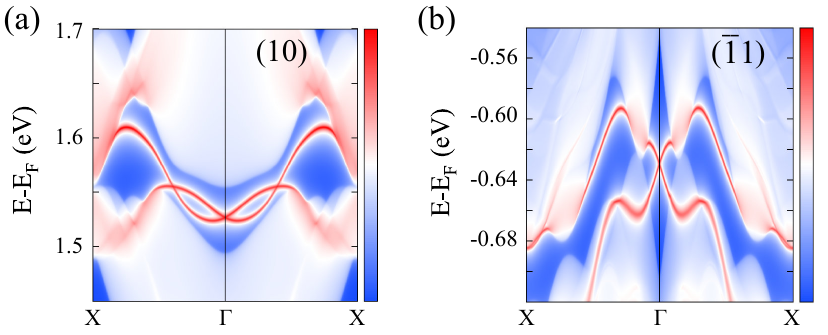}
	\caption{(a, b) Topological nontrivial helical edge states of Pythagorean iodinene along the (10) and ($\overline{1}$1) edge respectively.}
	\label{edgeStatePytha}
\end{figure}

Figure~\ref{edgeStatePytha} shows the topological helical edge states of Pythagorean iodinene in the nontrivial band gap.

\begin{figure}[h]
	\centering
	\includegraphics[scale=0.7]{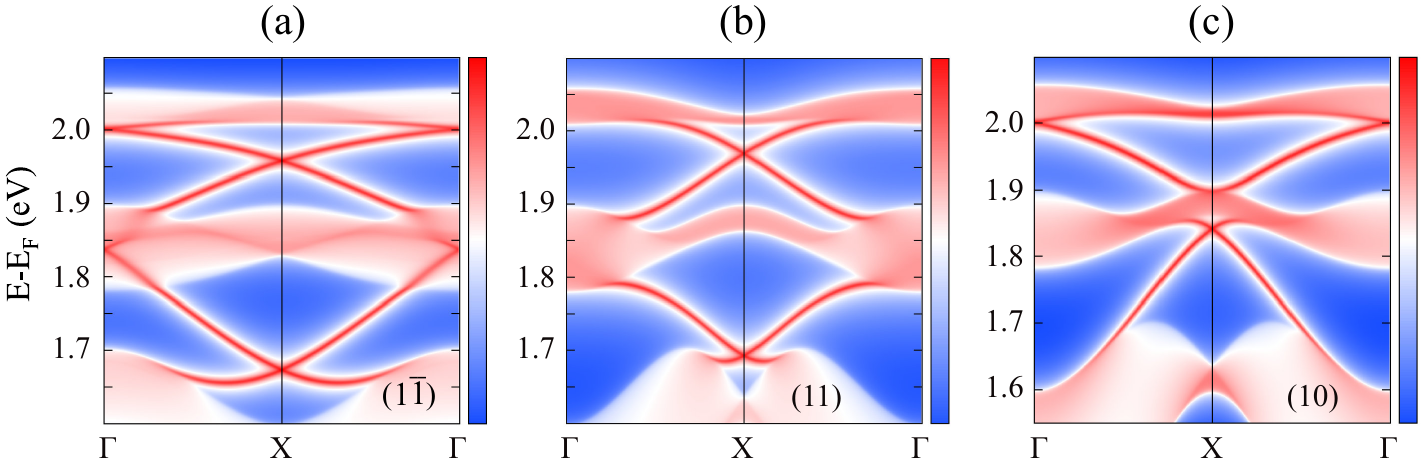}
	\caption{(a, b, c) Topological nontrivial helical edge states of GTH/diatomic-kagome iodinene along the (1$\overline{1}$), (11) and (10) edge respectively.}
	\label{edgeStateGTH}
\end{figure}

Figure~\ref{edgeStateGTH} shows the topological helical edge states of GTH/diatomic-kagome iodinene in the nontrivial band gap.

\begin{figure}[h]
	\centering
	\includegraphics[scale=0.7]{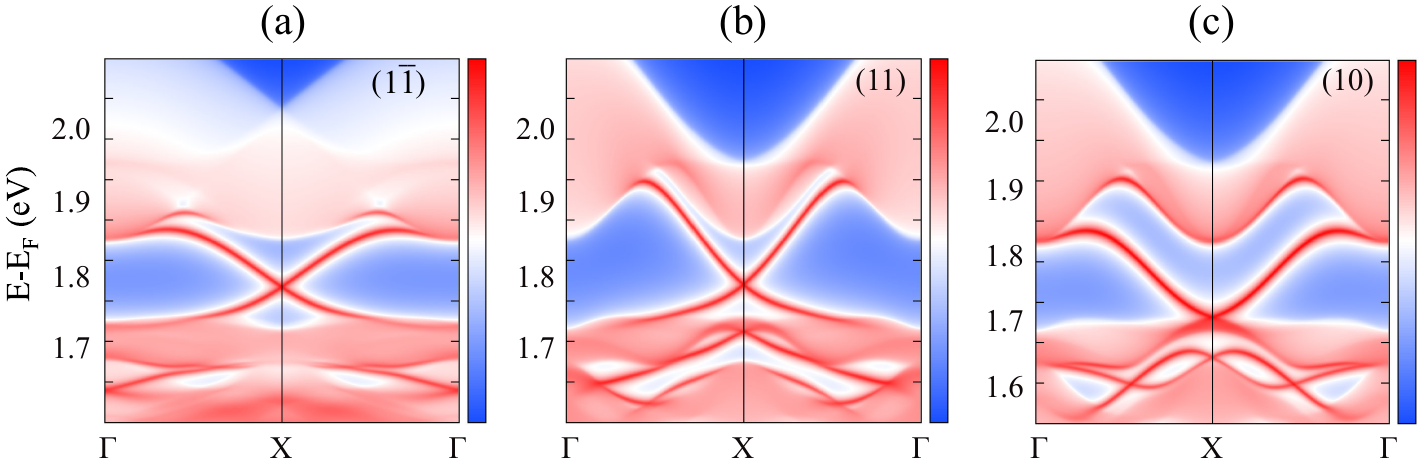}
	\caption{(a, b, c) Topological nontrivial helical edge states of GH iodinene along the (1$\overline{1}$), (11) and (10) edge respectively.}
	\label{edgeStateGH}
\end{figure}

Figure~\ref{edgeStateGH} shows the topological helical edge states of GH iodinene in the nontrivial band gap.

\subsubsection{Band structures and linear optical absorption spectra}
\begin{figure}[h]
	\centering
	\includegraphics[width=14cm]{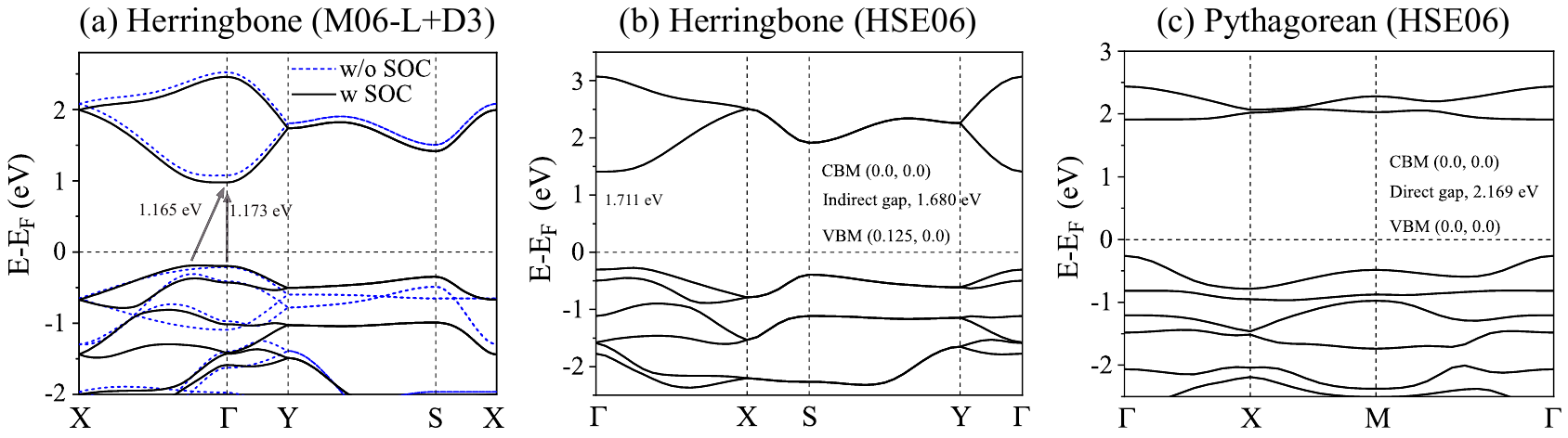}
	\caption{(a) Band structures of herringbone iodinene under M06-L+D3 scheme. Blue dashed lines and black lines are the band structures without and with spin-orbit coupling (SOC) respectively. Grey arrows indicate the band gap. (b,c) Band structures of herringbone and Pythagorean iodinene respectively under the HSE06 scheme with SOC.}
	\label{figS3}
\end{figure}

\begin{figure}[h]
	\centering
	\includegraphics[width=14cm]{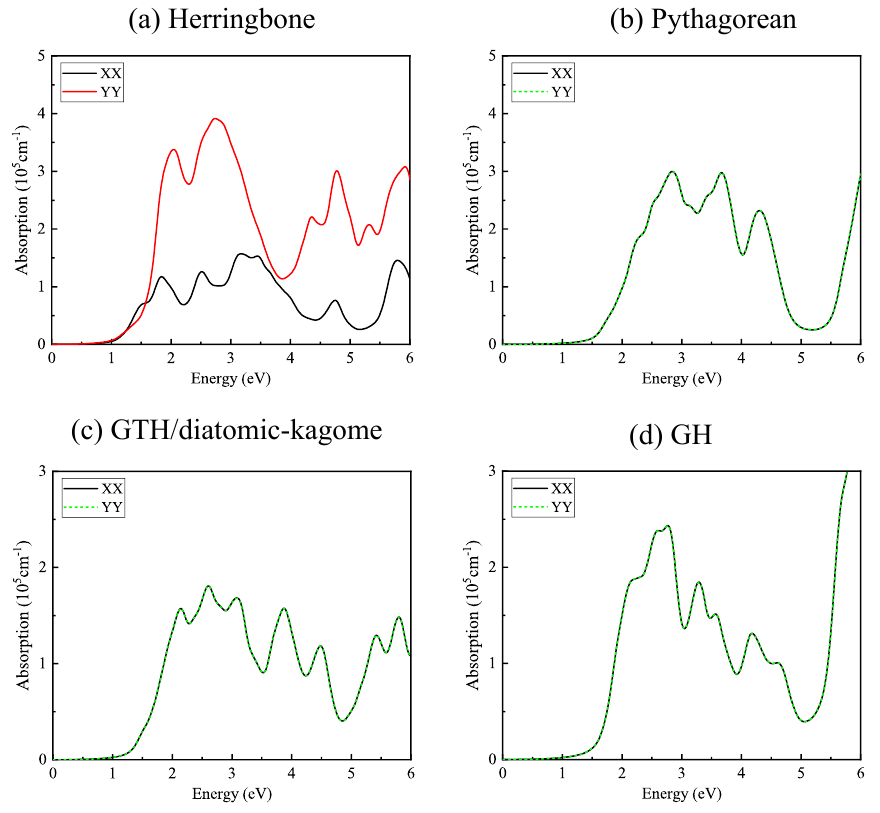}
	\caption{Computed in-plane optical absorption coefficient in units of $ 10^{5} cm^{-1} $ as a function of photon energy. "XX" and "YY" indicate the directions along two primitive vectors of a unit cell.}
	\label{figS4}
\end{figure}

Band structures are calculated using M06-L+D3 functional without and with SOC and HSE06 functional with SOC as shown in Fig.~\ref{figS3}. Linear optical absorption coefficients are calculated using M06-L+D3 functional with SOC as shown in Fig.~\ref{figS4}. 

\subsection{Results of iodinenes from the routine structure search method}

\begin{figure}[h]
	\centering
	\includegraphics[scale=0.55]{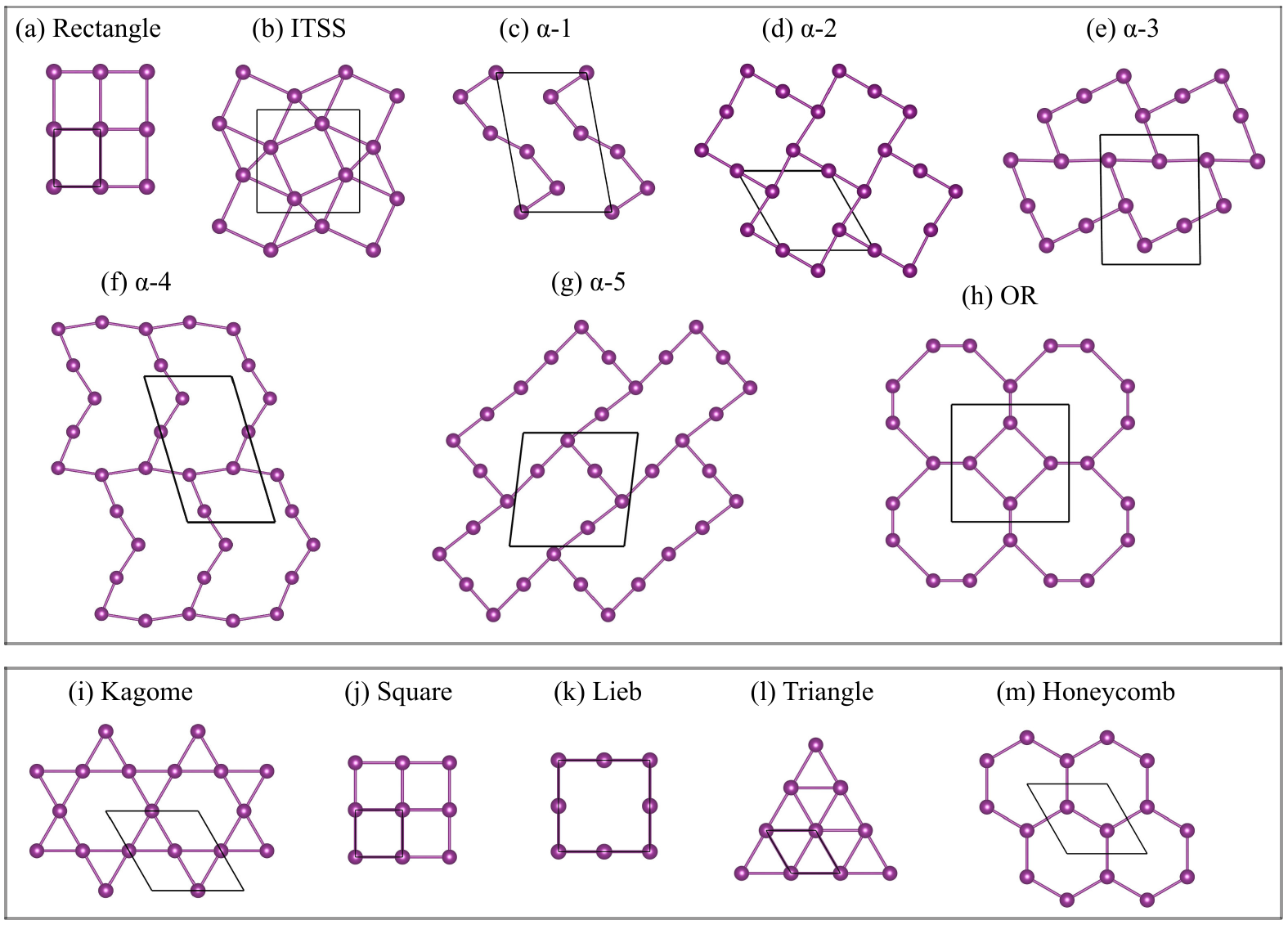}
	\caption{Crystal structures of monolayer iodinene from PSO searching (upper box) and hypothetical simple lattice (lower box).}
	\label{figS14}
\end{figure}

\begin{figure}[h]
	\centering
	\includegraphics[scale=0.8]{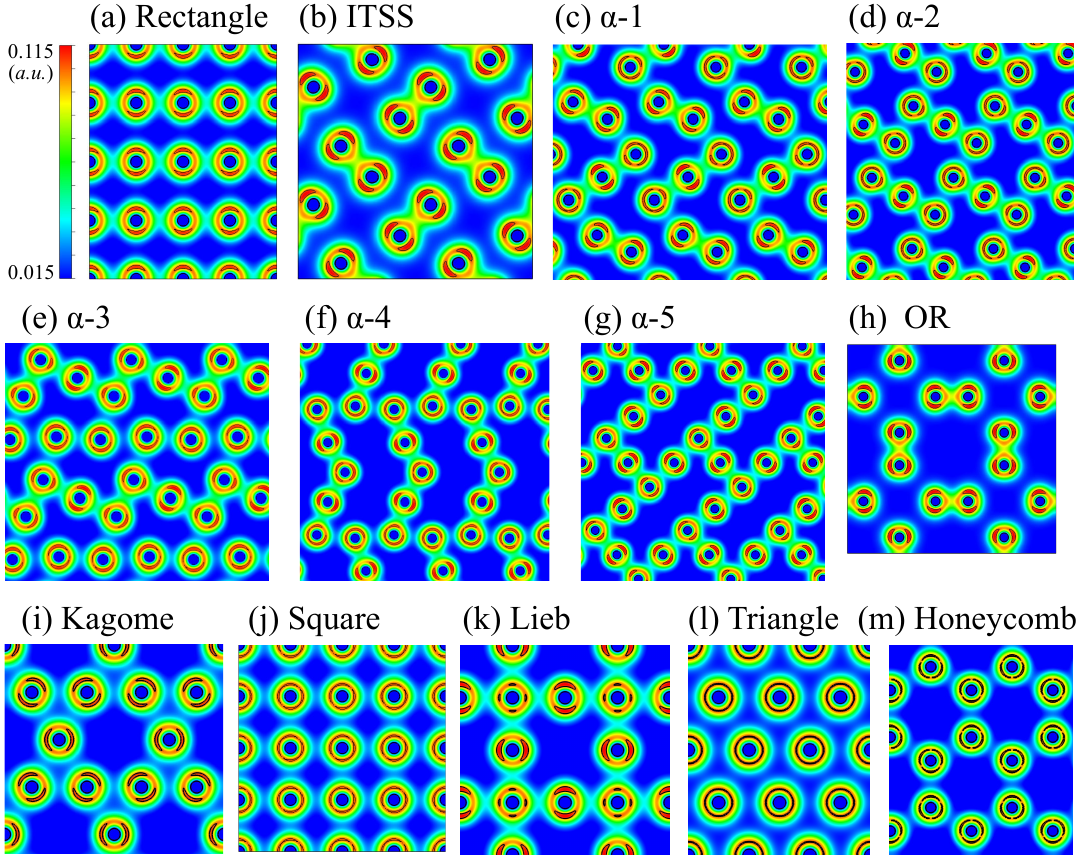}
	\caption{Valence electron charge density of iodinenes from (a-h) PSO searching and (i-m) hypothetical simple lattice. The unit of charge density is an atomic unit (e/bohr\textasciicircum{}3), and density inside the black crescent contour is > 0.1 a.u..}
	\label{figS15}
\end{figure}

\begin{figure}[h]
	\centering
	\includegraphics[scale=0.7]{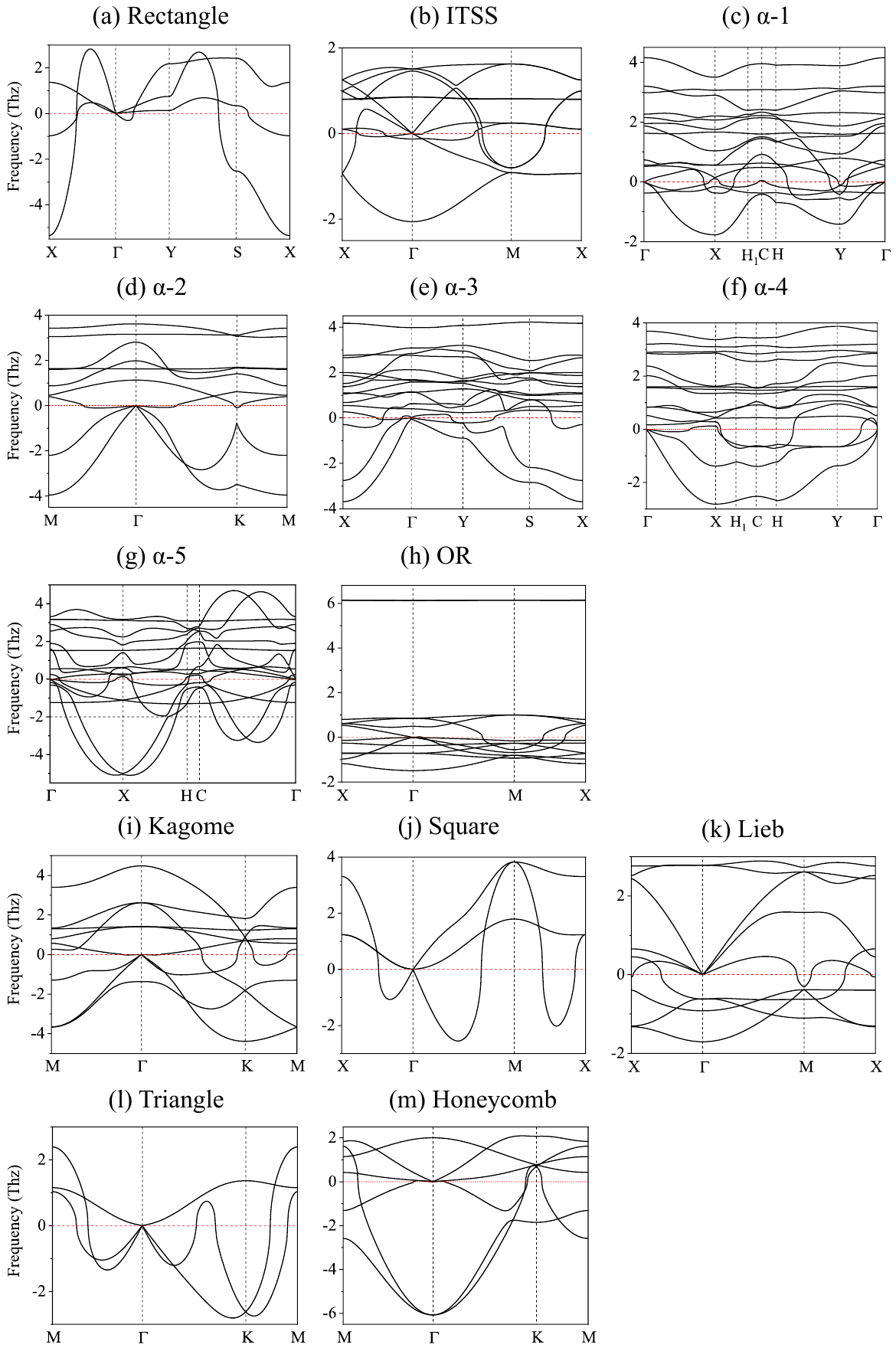}
	\caption{Phonon spectra of monolayer iodinenes from PSO searching (a-h) and hypothetical simple lattice (i-m).}
	\label{figS16}
\end{figure}

To see the rationality of constructing iodinenes using XBs, we performed a routine structure searching implemented in a Particle Swarm Optimization (PSO) algorithm. The particle swarm optimization (PSO) method implemented in CALYPSO\cite{calyspo2012} code is employed to search for low-energy 2D monolayer iodinenes. The population size is set to 30 and the number of generations is maintained to be 30 in our calculations. The number of iodine atoms in unit cells from 1 to 8 are considered. Apart from herringbone and Pythagorean pattern, structures searching results are shown in the upper panel of Fig.~\ref{figS14}. Details of crystal parameters and formation energies are listed in Table.~\ref{tableS2}. Among all the structures found, iodinenes constructed from diiodine molecules have the lowest energy configurations regardless of the calculation method used. Of course, M06L+D3 provides more accurate results. In Fig.~\ref{figS15}, we can see that only ITSS (Isosceles triangle snub square tiling pattern) and OR (Octagon rhombus tiling pattern) iodinenes are made from diiodines, but there are no XBs. Other structures look like traditional covalent crystals, thus all of them bear the dynamic instability as shown in the phonon spectra in Fig.~\ref{figS16}. Consequently, constructing the iodinene from diiodines combined with XBs is fruitful, especially considering that GH and GTH as low-energy configurations are omitted in PSO searching.

\end{document}